\documentstyle[12pt]{article}
\newlength{\abstractwidth}
\begin{document}
\thispagestyle{empty}
\pagestyle{empty}
\renewcommand{\thefootnote}{\fnsymbol{footnote}}
\renewcommand{\title}[1]{\begin{center}\large\bf #1\end{center}\par}
\renewcommand{\author}[1]{\vspace{2ex}{\normalsize\begin{center}
\setlength{\baselineskip}{3ex}#1\par\end{center}}}
\renewcommand{\thanks}[1]{\footnote{#1}} % Use this for footnotes
\renewcommand{\abstract}[1]{\vspace{2ex}\normalsize\begin{center}
\centerline{\bf Abstract}\par\vspace{2ex}\parbox{\abstractwidth}{{\small #1
\setlength{\baselineskip}{2.5ex}\par}}
\end{center}}
\newcommand{\starttext}{\normalsize
\pagestyle{plain}
\setlength{\baselineskip}{14pt}\par
\setcounter{footnote}{0}
\renewcommand{\thefootnote}{\arabic{footnote}}}
%%%%%%%%%%%%%%%%%%%%%%%%%%%%%%%%%%%%%%%%%%%%%%%%%%%%%%%%%%%%%%%
% Here is a replacement for the latex macro \section,         %
% which makes the section headings the correct size for W.S.  %
%%%%%%%%%%%%%%%%%%%%%%%%%%%%%%%%%%%%%%%%%%%%%%%%%%%%%%%%%%%%%%%
%\newcounter{mysection}
%\newcommand{\mysection}[1]{\stepcounter{mysection}
%\par\bigskip\noindent{\bf #1}\nopagebreak[4]\par\vskip .3cm}
%%%%%%%%%%%%%%%%%%%%%%%%%%%%%%%%%%%%%%%%%%%%%%%%%%%%%%%%%%%%%
%\newcommand{\mysectionstar}[1]{                             %  subsection macro
%\par\medskip\noindent{\it #1}\nopagebreak[4]\par\vskip .3cm}%
%%%%%%%%%%%%%%%%%%%%%%%%%%%%%%%%%%%%%%%%%%%%%%%%%%%%%%%%%%%%%
%Move the definition below past the style to get (c.s.e) equations
%\def\theequation{\themysection.\arabic{equation}}
%%%%%%%%%%%%%%%%%%%%%%%%%%%%%%%%%%%%%%%%%%%%%%%%%%%%%%%%%%%%%%%%

%%%%%%%%%%%%%%%%%%%%%%%%%%%%%%%%%%%%%%%%%%%%%%%%%%%%%%%%%%%%%%%
%%%%%%%%%%%%%%%%%%%   begin local macros %%%%%%%%%%%%%%%%%%%%%%
%%%%%%%%%%%%%%%%%%%%%%%%%%%%%%%%%%%%%%%%%%%%%%%%%%%%%%%%%%%%%%%

\def\ang{\,{\rm\AA}}
\def\flux{\,{\rm erg\,cm^{-2}\,arcsec^{-2}\,\AA^{-1}\,s^{-1}}}
\def\GeV{\,{\rm GeV}}
\def\TeV{\,{\rm TeV}}
\def\gev{\,{\rm GeV}}
\def\keV{\,{\rm keV}}
\def\MeV{\,{\rm MeV}}
\def\sec{\,{\rm sec}}
\def\Gyr{\,{\rm Gyr}}
\def\yr{\,{\rm yr}}
\def\rcm{\,{\rm cm}}
\def\pc{\,{\rm pc}}
\def\kpc{\,{\rm kpc}}
\def\Mpc{\,{\rm Mpc}}
\def\mpc{\,{\rm Mpc}}
\def\eV{{\,\rm eV}}
\def\ev{{\,\rm eV}}
\def\erg{{\,\rm erg}}
\def\cmm2{{\,\rm cm^{-2}}}
\def\cm2{{\,{\rm cm}^2}}
\def\cmm3{{\,{\rm cm}^{-3}}}
\def\gcmm3{{\,{\rm g\,cm^{-3}}}}
\def\kms{\,{\rm km\,s^{-1}}}
\def\HO{{100h\,{\rm km\,sec^{-1}\,Mpc^{-1}}}}
\def\mpl{{m_{\rm Pl}}}
\def\mpp{{m_{\rm Pl,0}}}
\def\trh{T_{\rm RH}}
\def\g{\tilde g}
\def\R{{\cal R}}
\def\km{\rm \,km} 
\def\yrs{\rm \,yrs} 
\def\trh{T_{\rm RH}} 

\def\baselinestretch{1.4}
\def\VEV#1{\left\langle #1\right\rangle}
\def\la{\mathrel{\mathpalette\fun <}}
\def\ga{\mathrel{\mathpalette\fun >}}
\def\fun#1#2{\lower3.6pt\vbox{\baselineskip0pt\lineskip.9pt
  \ialign{$\mathsurround=0pt#1\hfil##\hfil$\crcr#2\crcr\sim\crcr}}}

%%%%%%%%%%%%%%%%%%%%%%%%%%%%%%%%%%%%%%%%%%%%%%%%%%%%%%%%%%%%%%%%%%%%%
%%%%%%%%%%%%        end local macros      %%%%%%%%%%%%%%%%%%%%%%%%%%%
%%%%%%%%%%%%%%%%%%%%%%%%%%%%%%%%%%%%%%%%%%%%%%%%%%%%%%%%%%%%%%%%%%%%%

\title{COSMOLOGY:  STANDARD AND INFLATIONARY
\thanks{Supported in part by
the DOE (at Chicago and Fermilab) and by the NASA through
grant NAG 5-2788 (at Fermilab).}
}

\author{Michael S. Turner\\
{\small\it Departments of Physics and Astronomy \& Astrophysics,\\
Enrico Fermi Institute, The University of Chicago, Chicago, IL~~60637-1433}\\
\vspace{.1in}
{\it NASA/Fermilab Astrophysics Center,\\
Fermi National Accelerator Laboratory, Batavia, IL~~60510-0500}
}

\date{}
\abstract{\small
In these lectures I review the standard
hot big-bang cosmology, emphasizing its successes, its shortcomings,
and its major challenges---developing a detailed
understanding of the formation of structure in the 
Universe and identifying the constituents
of the ubiquitous dark matter.  I then discuss the motivations
for---and the fundamentals 
of---inflationary cosmology, particularly emphasizing the 
quantum origin of metric (density and gravity-wave) perturbations.
Inflation addresses the shortcomings of the standard cosmology,
specifies the nature of the dark matter,
and provides the ``initial data'' for structure formation.
I conclude by addressing the implications of inflation for structure formation
and discussing the different versions of cold dark matter.
The flood of data---from the Heavens and from Earth---should
in the next decade test inflation and discriminate between
the different cold dark matter models.}

\starttext
%%%%%%%%%%%%%%%%%%%%%%%%%%%%%%%%%%%%%%%%%%%%%%%%%%%%%%%%%%%%%%%%%%%%%%%%%%%
% Here comes examples of how to use the section and subsection macros.
% Note the extra space, "\," in the subsection, because of the italics.
% Also note that there is no counter for the subsection number.
%(You can put one in if you want)
%%%%%%%%%%%%%%%%%%%%%%%%%%%%%%%%%%%%%%%%%%%%%%%%%%%%%%%%%%%%%%%%%%%%%%%%%%%

\section{Hot Big Bang:  Successes and Challenges}
\subsection{Successes} 
 
The hot big-bang model, more properly the Friedmann-Robertson-Walker (FRW)
cosmology or standard cosmology, is spectacularly successful: 
In short, it provides a reliable and tested accounting 
of the history of the Universe from about $0.01\sec$ 
after the bang until today, some 15 billion years later. 
The primary pieces of evidence that support the model 
are:  (1) The expansion of the Universe; 
(2) The cosmic background radiation (CBR); (3) The
primordial abundances of the light elements D, $^3$He, 
$^4$He, and $^7$Li \cite{standard};  and (4) The existence
of small variations in the temperature of the CBR measured
in different directions (of order $30\mu$K on angular
scales from $0.5^\circ$ to $90^\circ$).
 
\subsubsection{The expansion}

Although the precise value of the Hubble constant
is not known to better than a factor of two, $H_0 =100h\,\km 
\sec^{-1}\Mpc^{-1}$ with $h=0.4-0.9$, there is little doubt
that the expansion obeys the ``Hubble law'' out to red shifts 
approaching unity \cite{h50,mould};
see Fig.~1.  As is well appreciated,
the fundamental difficulty in determining the Hubble constant 
is the calibration of the cosmic-distance scale as ``standard
candles'' are required \cite{distance1,distance2}.  The detection
of Cepheid variable stars in an Virgo Cluster galaxy (M101)
with the Hubble Space Telescope \cite{freedmanetal} was a giant
step toward an accurate determination of $H_0$, and the issue
could well be settled within five years.

\begin{figure}
\vspace{4.0in}
\caption[hubble]{Hubble diagram (from \cite{mould}).
The deviation from a linear relationship around $40\Mpc$
is due to peculiar velocities.}
\end{figure}
 
The Hubble law allows one to 
infer the distance to an object from its red shift $z$: 
$d = zH_0^{-1}\simeq 3000z\,h^{-1}\Mpc$ (for $z\ll 1$, 
the galaxy's recessional velocity $v\simeq zc$), and hence ``maps 
of the Universe'' constructed from galaxy positions and red shifts
are referred to as red-shift surveys.  Ordinary galaxies and clusters of
galaxies are seen out to red shifts of order unity; more 
unusual and rarer objects, such as radio galaxies and quasars, 
are seen out to red shifts of almost five (the current 
record holder is a quasar with red shift 4.9).  Thus, we can
probe the Universe with visible light to within a few billion years
of the big bang. 
 
\subsubsection{The cosmic background radiation}

The spectrum of the cosmic background radiation (CBR) is consistent
that of a black body at temperature 2.73 K over more than three
decades in wavelength ($\lambda \sim 0.03\rcm - 100\rcm$); see 
Fig.~2.  The most accurate measurement of the temperature
and spectrum is that by the FIRAS instrument on the
COBE satellite which determined its temperature to be
$2.726\pm 0.005\,$K \cite{FIRAS}.  It is difficult to
come up with a process other
than an early hot and dense phase in the history 
of the Universe that would lead to such a precise 
black body \cite{dnsnature}.  According to the standard cosmology,
the surface of last scattering for the CBR is 
the Universe at a red shift of about $1100$ and 
an age of about $180,000\,(\Omega_0 h^2)^{-1/2}\yrs$. 
It is possible that the Universe became ionized
again after this epoch, or due to energy injection
never recombined; in this case the last-scattering surface
is even ``closer,'' $z_{\rm LSS} \simeq 10[\Omega_Bh/\sqrt{\Omega_0}]^{-2/3}$.

\begin{figure}
\vspace{7.25in}
\caption[spectrum]{(a)  CBR spectrum as measured by
the FIRAS on COBE; (b) Summary of other CBR temperature measurements.
(Figure courtesy of G. Smoot.)}
\end{figure}
 
The temperature of the CBR is very uniform across the sky, to better than
a part in $10^4$ on angular scales from arcminutes
to 90 degrees; see Fig.~3.  Three forms of temperature anisotropy---two
spatial and one temporal---have now been detected: 
(1) A dipole anisotropy of about a part in 
$10^3$, generally believed to be due to the motion 
of galaxy relative to the cosmic rest frame, at a speed 
of about $620\km \sec^{-1}$ \cite{dipole}; (2) A yearly modulation in
the temperature in a given direction on the sky of 
about a part in $10^4$, due to our orbital motion 
around the sun at $30\km \sec^{-1}$, see Fig.~4 \cite{yearly}; and (3)
The temperature anisotropies detected by the Differential 
Microwave Radiometer (DMR) on the Cosmic
Background Explorer (COBE) satellite \cite{DMR} and
more than ten other experiments \cite{WSS}.

COBE has made the most precise measurement of CBR anisotropy,
$\langle (\Delta T /T)^2 \rangle_{10^\circ}^{1/2} =
1.1 \pm 0.1 \times 10^{-5}$ (the {\it rms} temperature fluctuation
averaged over the entire sky as measured by a beam of width $10^\circ$).
Other ground-based and balloon-borne instruments have now
measured CBR anisotropy on angular scales from about $0.5^\circ$ to $30^\circ$.
The CBR anisotropy provides strong evidence for primeval
density inhomogeneities of the same magnitude, which amplified by gravity,
grew into the structures that we see today:  galaxies, clusters of galaxies,
superclusters, voids, walls, and so on.  Moreover, CBR
anisotropy measurements are beginning to map out the inhomogeneity
on scales from about $100\Mpc$ to $10^4\Mpc$.

\begin{figure}
\vspace{3.5in}
\caption[anisotropy]{Summary of current measurements of CBR
anisotropy in terms of a spherical-harmonic decomposition,
$C_l \equiv \langle |a_{lm}|^2\rangle$.
The rms temperature fluctuation measured between two
points separated by an angle $\theta$
is roughly given by:  $(\delta T/T)_{\theta} \simeq
\sqrt{l(l+1)C_l}$ with $l \simeq 200^\circ /\theta$.
The curves are the cold dark matter predictions,
normalized to the COBE detection, for Hubble constants
of $50\kms\Mpc$ (solid) and $35\kms\Mpc^{-1}$ (broken).
(Figure courtesy of M.~White.)}
\end{figure}

\begin{figure}
\vspace{4in}
\caption[earth]{Yearly modulation of the CBR temperature---the
earth really orbits the sun(!) (from \cite{yearly}).}
\end{figure}

\subsubsection{Primordial nucleosynthesis}

Last, but certainly not least, there are the abundance
of the light elements.  According to the standard 
cosmology, when the age of the Universe was measured 
in seconds, the temperatures were of order MeV, and 
the conditions were right for nuclear reactions
which ultimately led to the synthesis of significant amounts
of D, $^3$He, $^4$He, and $^7$Li.
The yields of primordial nucleosynthesis depend 
upon the baryon density, quantified as the baryon-to-photon 
ratio $\eta$, and the number of very light ($\la \MeV$) 
particle species, often quantified as the equivalent number of 
light neutrino species, $N_\nu$.  The predictions for 
the primordial abundances of all four light elements 
agree with their measured abundances provided that 
$2.5\times 10^{-10} \la \eta \la 6\times 10^{-10}$ and
$N_\nu \la 3.9$; see Fig.~5 \cite{walkeretal,cst,kk,cst1}.

\begin{figure}
\vspace{7in}
\caption[bbn]{Predicted light-element abundances
including $2\sigma$ theoretical uncertainties (from \cite{cst}).
The inferred primordial abundances and concordance regions are indicated.}
\end{figure}

Accepting the success of the standard model of nucleosynthesis, 
our precise knowledge of the 
present temperature of the Universe allows us to 
convert $\eta$ to a mass density, and by dividing 
by the critical density, $\rho_{\rm crit} \simeq 
1.88h^2 \times 10^{-29}\gcmm3$, to the fraction 
of critical density contributed by ordinary matter: 
\begin{equation} 
0.009 \la \Omega_B h^2 \la 0.022;\qquad\Rightarrow
\ \   0.01 \la \Omega_B \la 0.15;
\end{equation} 
this is the most accurate determination of the baryon 
density.  Note, the uncertainty in the value of the 
Hubble constant leads to most of the uncertainty in $\Omega_B$.
 
The nucleosynthesis bound to $N_\nu$, and more generally 
to the number of light degrees of freedom in thermal 
equilibrium at the epoch of nucleosynthesis, is consistent 
with precision measurements of the properties of the $Z^0$ 
boson, which give $N_\nu = 3.0 \pm 0.02$; further, the
cosmological bound predates 
these accelerator measurements!  The nucleosynthesis
bound provides a stringent limit to the existence 
of new, light particles (even beyond neutrinos), and 
even provides a bound to the mass the tau neutrino, 
excluding a long-lived tau-neutrino of mass between $0.5\MeV$ and
$30\MeV$ \cite{tau,dgt}.  Primordial nucleosynthesis provides
a beautiful illustration of the powers of the Heavenly 
Laboratory, though it is outside the focus of these lectures. 
 
The remarkable success of primordial nucleosynthesis gives 
us confidence that the standard cosmology provides 
an accurate accounting of the Universe at least as 
early as $0.01\sec$ after the bang, when the temperature 
was about $10\MeV$. 
 
\subsubsection{Et cetera---and the age crisis?}

There are additional lines of reasoning and evidence
that support the standard cosmology \cite{dnsnature}.
I mention two:
the age of the Universe and structure formation.
I will discuss the basics of structure formation a bit
later; for now it suffices to say that the standard
cosmology provides a basic framework for understanding
the formation of structure---amplification of small
primeval density inhomogeneities through
gravitational instability.
Here I focus on the age of the Universe.

The expansion age of the Universe---time back to 
zero size---depends upon the present expansion rate, 
energy content, and equation of state:
$t_{\rm exp} = f(\rho ,p)H_0^{-1}\simeq 
9.8h^{-1} f(\rho ,p)\Gyr$.  For a matter-dominated Universe,
$f$ is between 1 and 2/3 (for $\Omega_0$ between 0 and 1),
so that the expansion age is somewhere between $7\Gyr$ and 
$20\Gyr$.  There are other independent measures of the 
age of the Universe, e.g., based upon long-lived radioisotopes, 
the oldest stars, and the cooling of white dwarfs.  These 
``ages,'' ranging from 13 to 18 Gyr, span the
same interval(!) \cite{age}.  This wasn't always
the case; as late as the early 1950's it was believe that 
the Hubble constant was $500\km \sec^{-1} \Mpc^{-1}$, implying 
an expansion age of at most $2\Gyr$---less than the age of the earth. 
This discrepancy was an important motivation for the steady-state cosmology.

While there is {\it general} agreement between the expansion age
and other determinations of the age of the Universe, some
cosmologists are worried that cosmology is on the verge
of another age crisis \cite{distance2}.  Let me explain, while Sandage and
a few others continue to obtain values for the Hubble constant
around $50\kms \Mpc^{-1}$ \cite{h50}, a variety of different techniques
seem to be converging on a value around $80\pm 10\kms \Mpc^{-1}$
\cite{distance2}.  If $H_0 =80\kms\Mpc^{-1}$, then $t_{\rm exp} =
12f(\rho ,p) \Gyr$, and for $\Omega_0 =1$,
$t_{\rm exp} =8\Gyr$, which is clearly inconsistent
with other measures of the age.  {\it If} $H_0=80\kms\Mpc^{-1}$,
one is almost forced to consider the radical alternative of
a cosmological constant.  For example, even with
$\Omega_0=0.2$, $f\simeq 0.85$, corresponding to
$t_{\rm exp} \simeq 10\Gyr$; on the other hand,
for a flat Universe with $\Omega_\Lambda
=0.7$, $f\simeq 1$ and the expansion age $t_{\rm exp}
\simeq 12\Gyr$.  As I shall discuss later, structure formation
provides another motivation for a cosmological constant.
As mentioned earlier, the detection of Cepheid variables
in Virgo \cite{freedmanetal} is a giant step toward an accurate determination
of $H_0$, and it seems likely that the issue may be settled soon.

\subsection {Basics of the Big Bang Model} 
 
The standard cosmology is based upon the maximally spatially symmetric 
Robertson-Walker line element 
\begin{equation} 
ds^2 = dt^2 -R(t)^2\left[ {dr^2\over 1-kr^2} +r^2 
        (d\theta^2 + \sin^2\theta\,d\phi^2 ) \right]; 
\end{equation} 
where $R(t)$ is the cosmic-scale factor, $R_{\rm curv}\equiv 
R(t)|k|^{-1/2}$ is the curvature radius, and $k/|k| = -1, 
0, 1$ is the curvature signature.  All three models are 
without boundary:  the positively curved model is finite 
and ``curves'' back on itself; the negatively curved 
and flat models are infinite in extent (though finite 
versions of both can be constructed by imposing a 
periodic structure:  identifying all points in 
space with a fundamental cube).  The Robertson-Walker 
metric embodies the observed isotropy and homogeneity of 
the Universe.  It is interesting to note
that this form of the line element was originally introduced 
for sake of mathematical simplicity; we now know that 
it is well justified at early times or today on large
scales ($\gg 10\Mpc$), at least within our Hubble volume.
 
The coordinates, $r$, $\theta$, and $\phi$, are referred 
to as comoving coordinates:  A particle at rest in these 
coordinates remains at rest, i.e., constant $r$, $\theta$, 
and $\phi$.  A freely moving particle eventually comes
to rest these coordinates, as its momentum is red shifted
by the expansion, $p \propto R^{-1}$.
Motion with respect to the comoving coordinates (or cosmic
rest frame) is referred to as peculiar velocity; unless
``supported'' by the inhomogeneous distribution of matter
peculiar velocities decay away as $R^{-1}$.  Thus the
measurement of peculiar velocities, which is not easy
as it requires independent measures of both the distance
and velocity of an object, can be used to probe the
distribution of mass in the Universe.

Physical separations (i.e.,
measured by meter sticks) between freely moving particles 
scale as $R(t)$; or said another way the physical separation 
between two points is simply $R(t)$ times the coordinate 
separation.  The momenta of freely propagating particles 
decrease, or ``red shift,'' as $R(t)^{-1}$, and thus the
wavelength of a photon stretches as $R(t)$, which is 
the origin of the cosmological red shift.  The red shift 
suffered by a photon emitted from a distant galaxy 
$1+z = R_0/R(t)$; that is, a galaxy whose light is 
red shifted by $1+z$, emitted that light when the Universe 
was a factor of $(1+z)^{-1}$ smaller.  Thus, when the 
light from the most distant quasar yet seen ($z=4.9$) was emitted 
the Universe was a factor of almost six smaller; when 
CBR photons last scattered the Universe was about $1100$ times smaller.
 
\subsubsection{Friedmann equation and the First Law}
 
The evolution of the cosmic-scale factor is governed 
by the Friedmann equation 
\begin{equation} 
H^2 \equiv \left({\dot R \over R}\right)^2 = 
        {8\pi G \rho_{\rm tot} \over 3} - {k\over R^2};
\end{equation} 
where $\rho_{\rm tot}$ is the total energy density of the
Universe, matter, radiation, vacuum energy, and so on.
A cosmological constant is often written as an additional
term ($=\Lambda /3$) on the rhs; I will choose to
treat it as a constant energy density (``vacuum-energy density''), where
$\rho_{\rm vac} = \Lambda /8\pi G$.  (My convention in this regard
is not universal.)  The evolution of the energy 
density of the Universe is governed by
\begin{equation} 
d(\rho R^3) = -pdR^3; 
\end{equation} 
which is the First Law of Thermodynamics for
a fluid in the expanding Universe.
(In the case that the stress energy of the Universe is comprised
of several, noninteracting components, this relation applies
to each separately; e.g., to the matter and radiation separately
today.)  For $p=\rho /3$, ultra-relativistic matter,
$\rho \propto R^{-4}$; for $p=0$, very nonrelativistic 
matter, $\rho \propto R^{-3}$; and for $p=-\rho$, vacuum 
energy, $\rho = \,$const.  If the rhs of the Friedmann 
equation is dominated by a fluid
with equation of state $p = \gamma \rho$, it follows 
that $\rho \propto R^{-3(1+\gamma )}$
and $R\propto t^{2/3(1+\gamma )}$. 
 
We can use the Friedmann equation to relate the 
curvature of the Universe to the energy density and 
expansion rate: 
\begin{equation} 
{k /R^2 \over H^2} = \Omega -1;\qquad 
\Omega = {\rho_{\rm tot}\over \rho_{\rm crit}};
\end{equation} 
and the critical density today $\rho_{\rm crit}
= 3H^2 /8\pi G = 1.88h^2\gcmm3 \simeq 1.05\times 10^{4}
\eV \cmm3$.  There is a one to one correspondence
between $\Omega$ and the spatial curvature of the Universe: 
positively curved, $\Omega_0 >1$; negatively curved, $\Omega_0
<1$; and flat ($\Omega_0 = 1$).  Further, the ``fate of the
Universe'' is determined by the curvature:  model universes 
with $k\le 0$ expand forever, while those with $k>0$ necessarily 
recollapse.  The curvature radius of the Universe is related 
to the Hubble radius and $\Omega$ by 
\begin{equation} 
R_{\rm curv} = {H^{-1}\over |\Omega -1|^{1/2}}. 
\end{equation} 
In physical terms, the curvature radius sets the scale for
the size of spatial separations where
the effects of curved space become 
``pronounced.''  And in the case of the positively curved
model it is just the radius of the 3-sphere.
 
The energy content of the Universe consists of matter 
and radiation (today, photons and neutrinos).  Since 
the photon temperature is accurately known,
$T_0=2.73\pm 0.01\,$K, the
fraction of critical density contributed by radiation 
is also accurately known:  $\Omega_{\rm rad}h^2 = 4.18 \times
10^{-5}$.  The matter content is another matter. 
 
\subsubsection {A short diversion concerning the present mass density}
 
The matter density today, i.e., the value of $\Omega_0$,
is not nearly so well
known \cite{dm}.  Stars contribute much less than 1\% of critical density;
based upon nucleosynthesis, we can infer that baryons 
contribute between 1\% and 15\% of critical.  The
dynamics of various systems allow astronomers to 
infer their gravitational mass.  With their telescopes
they measure the amount of light, and form a mass-to-light
ratio.  Multiplying this by the measured luminosity
density of the Universe gives a determination of the
mass density.  (The critical mass-to-light ratio is
$1200h\,M_\odot /{\cal L}_\odot$.)

The motions of stars and gas clouds in spiral galaxies indicate that
most of the mass of spiral galaxies exists in 
the form of dark (i.e., no detectable radiation), 
extended halos, whose full extent 
is still not known.  Many cite the
flat rotation curves of spiral galaxies,
which indicate that the halo density decreases
as $r^{-2}$, as the best evidence
that most of the matter in the Universe is dark.
Taking the mass-to-light ratio inferred for
spiral galaxies to be typical of the Universe as a
whole and remembering that the full extent of
the dark matter halos is not known,
one infers $\Omega_{\rm halo} \ga 0.03 - 0.1$ \cite{nick}.

The masses of clusters of galaxies have been determined
by applying the virial theorem to the motions of
member galaxies or to the hot gas that fills the
intracluster medium, and by the analyzing
(weak) gravitational lensing of very distant
galaxies by clusters.  These mass estimates
too indicate the presence of large amounts of 
dark matter; when more than one method is applied
to the same cluster the mass estimates are consistent.
Taking cluster mass-to-light ratios
to be typical of the Universe as a whole, in spite
of the fact that only about 1 in 10 galaxies resides
in a cluster, one infers $\Omega_{\rm cluster} \sim 0.2-0.4$.

Another interesting fact has been learned from x-ray
observations of clusters:  the ratio of baryons in the hot intracluster
gas to the total cluster mass, $M_{\rm gas}/M_{\rm tot}
\simeq (0.04 -0.08)h^{-3/2}$ \cite{gasratio}.  Since the
gas mass is much greater than the mass in the visible galaxies,
this ratio provides an estimate of the cluster baryon
fraction, {\it provided that most of the baryons reside in
the hot gas or in galaxies,} and suggests that the bulk of
matter in clusters is in a form other than baryons!

Not one of these methods is wholly satisfactory:
Rotation curves of spiral galaxies are still ``flat'' 
at the last measured points, indicating that the 
mass is still increasing; likewise, cluster virial 
mass estimates are insensitive to material that lies 
beyond the region occupied by the visible galaxies---and 
moreover, only about one galaxy in ten resides in 
a cluster.   What one would like is a measurement of the mass of a 
very big sample of the Universe, say a cube of 
$100h^{-1}\Mpc$ on a side, which contains tens of thousands  of galaxies.
 
Over the past five years or so progress 
has been made toward such a measurement.  It involves 
the peculiar motion of our own galaxy, at a speed of about 
$620\km\sec^{-1}$ in the general direction of
Hydra-Centaurus.  This motion is due to the lumpy
distribution of matter in our vicinity.  By using 
gravitational-perturbation theory (actually, not much more 
than Newtonian physics) and the distribution of galaxies in our vicinity
(as determined by the IRAS catalogue of infrared selected 
galaxies), one can infer the average mass density in 
a very large volume and thereby $\Omega_0$.

The basic physics behind the method is simple: the
net gravitational pull on our galaxy depends both upon
how inhomogeneous the distribution of galaxies is
and how much mass is associated with each galaxy;
by measuring the distribution of galaxies and our
peculiar velocity one can infer the ``mass per galaxy''
and $\Omega_0$.

The value that has been inferred is big(!)---close to unity---
and provides a very strong case that $\Omega_0$ is at
least 0.3 \cite{irasomega}.  Moreover,
the measured peculiar velocities of
other galaxies in this volume, more than thousand, 
have been used in a similar manner and indicate 
a similarly large value for $\Omega_0$ \cite{potent}.
While this technique is very powerful, 
it does have its drawbacks:  One has to make simple 
assumptions about how accurately 
mass is traced by light (the observed galaxies); 
one has to worry whether or not a significant portion 
of our galaxy's velocity is due to galaxies outside the 
IRAS sample---if so, this would lead to an overestimate 
of $\Omega_0$; and so on.  This technique is not only
very promising---but provides the ``correct'' answer
(in my opinion!).
 
The so-called classical kinematic tests---Hubble diagram, 
angle-red shift relation, galaxy count-red shift relation---can, 
in principle, provide a determination of $\Omega_0$ by determining
the deceleration parameter $q_0$ \cite{Sandage}.  However, all these methods
require standard candles, rulers, or galaxies, and for this reason
have proved inconclusive.  However, 
that hasn't discouraged everyone.  There are a number of efforts
to determine $q_0$ using
the galaxy number-count test \cite{lohspillar}, and two groups
are trying to measure $q_0$ by constructing
a Hubble diagram based upon Type Ia supernovae (out to
redshifts of 0.5 or more).
 
To summarize this aside on the mass density of the 
Universe: 
\begin{enumerate}

\item{}  Most of the matter is dark.
 
\item{}  Baryons provide between about 1\% and 15\% of the mass
density (allowing $0.4 <h< 1$; taking $h> 0.6$
the upper limit decreases to 6\%).
 
\item{}  There is a strong case that $\Omega_0 \ga
0.3$ (peculiar velocities); a convincing case that
$\Omega_0 \ga 0.2$ (cluster masses); and an airtight case that
$\Omega_0 \ga 0.1$ (flat rotation curves of spirals).

\item{}  Most of the baryons are dark (not in stars).
In clusters the bulk of the baryons are in hot gas.

\item{}  The evidence for nonbaryonic dark matter
continues to mount; e.g., the gap between $\Omega_B$ and $\Omega_0$
and the cluster baryon fraction.

\end{enumerate} 
 
The current prejudice---and certainly that of this author---is
a flat Universe ($\Omega_0 = 1$) with nonbaryonic
dark matter, $\Omega_X\sim 1\gg \Omega_B$.  However, I shall continue
to display the $\Omega_0$ dependence of important quantities.
 
\subsubsection{The early, radiation-dominated Universe} 
 
In any case, at present, matter outweighs radiation 
by a wide margin.  However, since the energy density 
in matter decreases as $R^{-3}$, and that 
in radiation as $R^{-4}$ (the extra factor due 
to the red shifting of the energy of relativistic 
particles), at early times the 
Universe was radiation dominated---indeed the calculations 
of primordial nucleosynthesis provide excellent evidence for this. 
Denoting the epoch of matter-radiation equality 
by subscript `EQ,' and using $T_0=2.73\,$K, it follows that
\begin{equation} 
R_{\rm EQ} = 4.18\times 10^{-5}\,(\Omega_0 h^2)^{-1};\qquad
T_{\rm EQ} = 5.62 (\Omega_0 h^2)\eV;
\end{equation}
\begin{equation}
t_{\rm EQ} = 4.17 \times 10^{10}(\Omega_0 h^2)^{-2}\sec .
\end{equation} 
At early times the expansion rate and age of the Universe were 
determined by the temperature of the Universe and 
the number of relativistic degrees of freedom: 
\begin{equation} 
\rho_{\rm rad} = g_*(T){\pi^2T^4 \over 30}; \qquad 
H\simeq 1.67g_*^{1/2} T^2 /\mpl ; 
\end{equation} 
\begin{equation} 
\Rightarrow R\propto t^{1/2}; \qquad 
t \simeq 2.42\times 10^{-6} g_*^{-1/2}(T/\GeV )^{-2}\,\sec ;
\end{equation} 
where $g_* (T)$ counts the number of ultra-relativistic 
degrees of freedom ($\approx$ the sum of the internal 
degrees of freedom of particle species much less massive 
than the temperature) and $\mpl \equiv G^{-1/2} = 1.22 
\times 10^{19}\GeV$ is the Planck mass.  For example, 
at the epoch of nucleosynthesis, $g_* = 10.75$ assuming 
three, light ($\ll \MeV$) neutrino species; taking into 
account all the species in the standard model, 
$g_* = 106.75$ at temperatures much greater than $300\GeV$; see Fig.~6.

\begin{figure}
\vspace{4in}
\caption[g*]{The total effective number of relativistic
degrees of freedom $g_*(T)$ in the standard model of particle physics
as a function of temperature.}
\end{figure}

A quantity of importance related to $g_*$ is the 
entropy density in relativistic particles, 
$$s= {\rho +p \over T} = {2\pi^2\over 45}g_* T^3 ,$$ 
and the entropy per comoving volume, 
$$S \ \  \propto\ \  R^3 s\ \  \propto\ \   g_*R^3T^3 .$$
By a wide margin most of the entropy 
in the Universe exists in the radiation bath. 
The entropy density is proportional
to the number density of relativistic particles. 
At present, the relativistic particle species 
are the photons and neutrinos, and the 
entropy density is a factor 
of 7.04 times the photon-number density:
$n_\gamma =413 \cmm3$ and $s=2905 \cmm3$.
 
In thermal equilibrium---which provides a good description 
of most of the history of the Universe---the entropy per comoving 
volume $S$ remains constant.  This fact is very useful. 
First, it implies that the temperature and scale 
factor are related by 
\begin{equation} 
T\propto g_*^{-1/3}R^{-1},
\end{equation} 
which for $g_*=\,$const leads to the familiar $T\propto R^{-1}$. 

Second, it provides a way of quantifying the net baryon 
number (or any other particle number) per comoving volume: 
\begin{equation} 
N_B \equiv R^3n_B = {n_B\over s} \simeq (4-7)\times 10^{-11}.
\end{equation} 
The baryon number of the Universe tells us two things:
(1) the entropy per particle in the Universe
is extremely high, about
$10^{10}$ or so compared to about $10^{-2}$ in the sun
and a few in the core of a newly formed neutron star.
(2)  The asymmetry between matter and antimatter is
very small, about $10^{-10}$, since at early times
quarks and antiquarks were roughly as abundant as
photons.  One of the great successes of particle
cosmology is baryogenesis, the idea that $B$, $C$, and
$CP$ violating interactions occurring out-of-equilibrium
early on allow the
Universe to develop a net baryon number of this magnitude
\cite{baryo}.

Finally, the constancy of the entropy per comoving
volume allows us to characterize the size of comoving
volume corresponding to our present Hubble volume in
a very physical way:  by the entropy it contains, 
\begin{equation} 
S_{U} = {4\pi\over 3}H_0^{-3}s \simeq 10^{90}.
\end{equation} 
 
\subsubsection{The earliest history}

The standard cosmology is tested back to times as early
as about 0.01 sec; it is only natural to ask how far back 
one can sensibly extrapolate.  Since the fundamental
particles of Nature are point-like quarks and leptons 
whose interactions are perturbatively weak at energies much greater
than $1\GeV$, one can imagine extrapolating
as far back as the epoch where general relativity
becomes suspect, i.e., where quantum gravitational 
effects are likely to be important:  the Planck epoch, 
$t\sim 10^{-43}\sec$ and $T\sim 10^{19}\GeV$.
Of course, at present, our firm understanding 
of the elementary particles and their interactions 
only extends to energies of the order of $100\GeV$, 
which corresponds to a time of the order of
$10^{-11}\sec$ or so.  We can be relatively certain 
that at a temperature of $100\MeV -200\MeV$ ($t\sim 10^{-5}\sec$)
there was a transition (likely a second-order
phase transition) from quark/gluon
plasma to very hot hadronic matter, and that some
kind of phase transition associated
with the symmetry breakdown of the electroweak theory 
took place at a temperature of the order of $300\GeV$
($t\sim 10^{-11}\sec$).
 
It is interesting to look at the progress that 
has taken place since Weinberg's classic text on 
cosmology was published in 1972 \cite{Weinberg}; at that time many
believed that the Universe had a limiting temperature 
of the order of several hundred $\MeV$, due to the
exponentially rising number of particle states,
and that one could not speculate about earlier
times.  Today, based upon our present knowledge of physics
and powerful mathematical tools (e.g., gauge 
theories, grand unified theories, and superstring
theory) we are able to make quantitative
speculations back to the Planck epoch---and
even earlier.  Of course, these speculations could be totally
wrong, based upon a false sense of confidence (arrogance?). 
As I shall discuss, inflation is one of these
well defined---and well motivated---speculations 
about the history of the Universe well after the 
Planck epoch, but well before primordial nucleosynthesis. 
 
\subsubsection{The matter and curvature dominated epochs} 
 
After the equivalence epoch, the matter density exceeds 
that of radiation.  During the matter-dominated epoch
the scale factor grows as $t^{2/3}$ and the age of the 
Universe is related to red shift by
\begin{equation} 
t = 2.06\times 10^{17} (\Omega_0 h^2)^{-1/2}(1+z)^{-3/2}\sec .
\end{equation} 
 
If $\Omega_0 < 1$, the matter-dominated epoch is
followed by a ``curvature-dominated'' epoch where the 
rhs of the Friedmann equation is dominated by the $|k|/R^2$ term.
When the Universe is curvature dominated it is said to
expand freely, no longer decelerating since the gravitational
effect of matter has become negligible:  $\ddot R\approx
0$ and $R\propto t$.
The epoch of curvature dominance
begins when the matter and curvature terms are equal: 
\begin{equation} 
R_{\rm CD} = {\Omega_0\over 1- \Omega_0} \longrightarrow \Omega_0;
        \qquad z_{\rm CD} = \Omega_0^{-1} - 2 \longrightarrow \Omega_0^{-1};
        \end{equation} 
where the limits shown are for $\Omega_0\rightarrow 0$.
By way of comparison, in a flat Universe with a cosmological 
constant, the Universe becomes ``vacuum dominated'' 
when $R=R_{\rm vac}$:
\begin{equation} 
R_{\rm vac} = \left( {\Omega_0 \over 1-\Omega_0} \right)^{1/3}
\longrightarrow \Omega_0^{1/3}; \qquad z_{\rm vac}
= \left( {1-\Omega_0 \over \Omega_0}\right)^{1/3} -1
\longrightarrow \Omega_0^{-1/3} .
\end{equation} 
For a given value of $\Omega_0$, the transition occurs
much more recently, which has important
implications for structure formation since small
density perturbations only grow during the matter-dominated era.

\subsubsection{One last thing:  horizons}

In spite of the fact that the Universe was
vanishingly small at early times, the rapid expansion
precluded causal contact from being established throughout.
Photons travel on null paths characterized
by $dr=dt/R(t)$; the physical distance that a photon
could have traveled since the bang until time $t$, the
distance to the horizon, is
\begin{eqnarray}
d_H(t) & = & R(t)\int_0^t {dt^\prime\over R(t^\prime )} \nonumber\\
       & = & t/(1-n) = nH^{-1}/(1-n) \qquad {\rm for}\ R(t)
        \propto t^n, \ \ n<1 .
\end{eqnarray}
Note, in the standard cosmology the distance to the
horizon is finite, and up to numerical factors,
equal to the age of the Universe or the Hubble radius, $H^{-1}$.
For this reason, I will use horizon and Hubble radius
interchangeably.\footnote{In inflationary models
the horizon and Hubble radius are not roughly equal
as the horizon distance grows exponentially relative
to the Hubble radius; in fact, at the end of inflation
they differ by $e^N$, where $N$ is the number of
e-folds of inflation.  However, I will slip and use
``horizon'' and ``Hubble radius'' interchangeably, though
I will always mean Hubble radius.}

An important quantity is
the entropy within a horizon volume:  $S_{\rm HOR}
\sim H^{-3}T^3$; during the radiation-dominated epoch
$H\sim T^2/\mpl$, so that
\begin{equation}
S_{\rm HOR} \sim \left( {\mpl\over T} \right)^3;
\end{equation}
from this we conclude that at early times the comoving
volume that encompasses all that we can see today
(characterized by an entropy of $10^{90}$) was comprised
of a very large number of causally disconnected regions.
 
\subsection {Two challenges:  dark matter and structure formation}

These two challenges are not unrelated:  a detailed understanding
of the formation of structure in the Universe necessarily
requires knowledge of the quantity and composition of
matter in the Universe.

We have every indication that the Universe at early
times, say $t\ll 300,000\yrs$, was very homogeneous;
however, today inhomogeneity (or structure) is ubiquitous:
stars ($\delta\rho /\rho \sim 10^{30}$),
galaxies ($\delta\rho /\rho \sim 10^{5}$),
clusters of galaxies ($\delta\rho /\rho \sim 10-10^{3}$),
superclusters, or ``clusters of clusters''
($\delta\rho /\rho \sim 1$), voids ($\delta\rho /\rho
\sim -1$), great walls, and so on.

For some 25 years the standard cosmology has provided
a general framework for understanding this:
Once the Universe becomes matter dominated (around 1000 yrs
after the bang) primeval density
inhomogeneities ($\delta\rho /\rho \sim 10^{-5}$)
are amplified by gravity and
grow into the structure we see today \cite{sf}.
The fact that a fluid of
self-gravitating particles is unstable to the growth of 
small inhomogeneities was first pointed out by Jeans
and is known as the Jeans instability.  The
existence of these inhomogeneities was confirmed in spectacular 
fashion by the COBE DMR discovery of CBR anisotropy.

At last, the basic picture
has been put on firm ground (whew!).
Now the challenge is to fill in the details---origin
of the density perturbations, precise evolution of
the structure, and so on.  As I shall emphasize,
such an understanding may well be within reach,
and offers a window on the early Universe.
 
\subsubsection{The general picture:  gravitational instability}

Let us begin by expanding the
perturbation to the matter density in plane waves
\begin{equation}
{\delta\rho_M ({\bf x}, t)\over\rho_M} = {1\over (2\pi )^3}\int d^3k\,
	\delta_k(t) e^{-i{\bf k}
	\cdot {\bf x}},
\end{equation}
where $\lambda = 2\pi /k$ is the comoving wavelength
of the perturbation and $\lambda_{\rm phys} =
R\lambda$ is the physical wavelength.
The comoving wavelengths of perturbations corresponding
to bright galaxies, clusters, and the present horizon scale
are respectively:  about $1\Mpc$, $10\Mpc$, and $3000h^{-1}\Mpc$,
where $1\Mpc \simeq 3.09\times 10^{24}\rcm
\simeq 1.56\times 10^{38}\GeV^{-1}$.

The growth of small matter inhomogeneities of wavelength
smaller than the Hubble scale ($\lambda_{\rm phys} \la
H^{-1}$) is governed by a Newtonian equation:
\begin{equation}
{\ddot\delta}_k + 2H{\dot\delta}_k +v_s^2k^2\delta_k /R^2
= 4\pi G\rho_M \delta_k ,
\end{equation}
where $v_s^2 = dp /d\rho_M$ is the square of the sound
speed.  Competition between the pressure term and
the gravity term on the rhs determine whether or
not pressure can counteract gravity:
Perturbations with wavenumber larger than the Jeans wavenumber,
$k_J^2 = 4\pi GR^2 \rho_M /v_s^2$, are Jeans stable
and just oscillate; perturbations with smaller
wavenumber are Jeans unstable and can grow.
For cold dark matter $v_s \simeq 0$ and all scales are
Jeans unstable; even for baryonic matter, after decoupling
$k_J$ corresponds to a baryon mass of only about $10^5M_\odot$.
All the scales of interest here are Jeans unstable
and we will ignore the pressure term.

Let us discuss solutions to this equation under
different circumstances.  First, consider the Jeans
problem, evolution of perturbations in a static fluid,
i.e., $H=0$.  In this case Jeans unstable
perturbations grow exponentially,
$\delta_k \propto \exp (t/\tau )$ where $\tau = 1/\sqrt{4G\pi\rho_M}$.
Next, consider the growth of Jeans unstable perturbations in a
matter-dominated Universe,
i.e., $H^2=8\pi G\rho_M/3$ and $R\propto t^{2/3}$.  Because
the expansion tends to ``pull particles away from
one another,'' the growth is only power law,
$\delta_k \propto t^{2/3}$; i.e., at the same rate as the
scale factor.  Finally, consider a radiation or curvature
dominated Universe, i.e., $8\pi G\rho_{\rm rad}/3$ or
$|k|/R^2$ much greater than $8\pi G\rho_M/3$.  In this case, the expansion
is so rapid that matter perturbations grow very slowly,
as $\ln R$ in radiation-dominated epoch, or not at all
$\delta_k =\,$const in the curvature-dominated epoch.

The growth of nonlinear perturbations is another matter;
once a perturbation reaches an overdensity of order unity
or larger it ``separates'' from the expansion---i.e., becomes
its own self-gravitating system and ceases to expand
any further.  In the process of virial
relaxation, its size decreases by a factor of two---density
increases by a factor of 8; thereafter, its density contrast
grows as $R^3$ since the average
matter density is decreasing as $R^{-3}$, though smaller
scales could become Jeans unstable and collapse further to form
smaller objects of higher density, stars, etc.

From this we learn that structure formation begins when
the Universe becomes matter dominated and ends when
it becomes curvature dominated (at least the growth
of linear perturbations).  The total growth available for linear
perturbations is $R_{\rm CD}/R_{\rm EQ} \simeq 2.4\times 10^4\,
\Omega_0^2h^2$; since nonlinear structures have evolved
by the present epoch, we can infer that primeval perturbations of
the order $(\delta\rho_M/\rho_M)_{\rm EQ} \sim 4\times 10^{-5}\,
(\Omega_0 h)^{-2}$ are required.  Note that in a low-density
Universe larger initial perturbations are necessary as there
is less time for growth (``the low
$\Omega_0$ squeeze'').  Further, in a baryon-dominated Universe
things are even more difficult as perturbations in the baryons
cannot begin to grow until after decoupling since matter is tightly
coupled to the radiation.  (In a flat, low-$\Omega_0$ model
with a cosmological constant the growth of linear
fluctuations continues until almost today since $z_\Lambda
\sim \Omega_0^{-1/3}$, and so the total growth factor
is about $2.4\times 10^4 (\Omega_0h^2)$.  We will return to
this model later.)

\subsubsection{CBR temperature fluctuations}

The existence of density inhomogeneities has another important
consequence:  fluctuations in the temperature of the CBR of
a similar amplitude \cite{dt/t}.  The temperature difference measured
between two points separated by a large angle ($\ga 1^\circ$)
arises due to a very simple physical effect:\footnote{Large
angles mean those larger than the angle
subtended by the horizon-scale at decoupling, $\theta
\sim H_{\rm DEC}^{-1}/H_0^{-1}\sim z_{\rm DEC}^{-1/2}\sim 1^\circ$.}
The difference in the gravitational potential between the two points on
the last-scattering surface, which in turn is related
to the density perturbation, determines
the temperature anisotropy on the angular scale subtended
by that length scale,
\begin{equation}
\left({\delta T \over T}\right)_\theta =
-\left({\delta\phi \over 3}\right)_\lambda \approx {1\over 2}
\left( {\delta\rho\over \rho}\right)_{{\rm HOR},\lambda};
\end{equation}
where the scale $\lambda \sim 100h^{-1}\Mpc(\theta /{\rm deg})$
subtends an angle $\theta$ on the last-scattering
surface.  This is known as the Sachs-Wolfe effect \cite{SW}.

The quantity $(\delta\rho /\rho)_{{\rm HOR},\lambda}$ is the
amplitude with which a density perturbation crosses
inside the horizon, i.e., when $R\lambda \sim H^{-1}$.  Since
the fluctuation in the gravitational potential
$\delta \phi \sim (R\lambda /H^{-1})^2(\delta\rho /\rho )$,
the horizon-crossing amplitude is equal to the
gravitational potential (or curvature) fluctuation.
The horizon-crossing amplitude $(\delta\rho /\rho )_{\rm HOR}$
has several nice features:  (i) during the matter-dominated era
the potential fluctuation on a given scale remains constant, and
thus the potential fluctuations at decoupling on scales
that crossed inside the horizon after matter-radiation
equality, corresponding to angular scales $\la 0.1^\circ$, are just
given by their horizon-crossing amplitude;
(ii) because of its relationship to $\delta\phi$
it provides a dimensionless,
geometrical measure of the size of the density perturbation
on a given scale, and its effect on the CBR; (iii) by specifying
perturbation amplitudes at horizon crossing one can effectively
avoid discussing the evolution of density perturbations
on scales larger than the horizon, where a Newtonian analysis
does not suffice and where gauge subtleties (associated
with general relativity) come into play; and finally (iv) the
density perturbations generated in inflationary models
are characterized by $(\delta\rho /\rho )_{\rm HOR}\simeq
\,$const.

On angular scales smaller than about $1^\circ$ two other
physical effects lead to CBR temperature fluctuations:
the motion of the last-scattering surface (Doppler) and
the intrinsic fluctuations in the local photon temperature.
These fluctuations are much more difficult to compute, and
depend on microphysics---the ionization history of the Universe
and the damping of perturbations in the photon-baryon fluid
due to photon streaming.  Not only are the Sachs-Wolfe fluctuations
simpler to compute, but they accurately mirror the
primeval fluctuations since at the epoch of decoupling
microphysics is restricted to angular scales less than about a degree.

In sum, on large angular scales the
Sachs-Wolfe effect dominates; on the scale of about
$1^\circ$ the total CBR fluctuation is about twice that due
to the Sachs-Wolfe effect; on smaller scales the Doppler and
intrinsic fluctuations dominate (see Fig.~3).  CBR temperature fluctuations
on scales smaller than about $0.1^\circ$ are severely
reduced by the smearing effect of the finite thickness of
last-scattering surface.  (For a beautiful exposition of how
CBR anisotropy arises see Ref.~\cite{wayne}.)

Details aside, in the context of the gravitational instability
scenario density perturbations of sufficient amplitude
to explain the observed structure lead to temperature
fluctuations in the CBR of characteristic size,
\begin{equation}
{\delta T\over T} \approx 10^{-5}\,(\Omega_0 h)^{-2}.
\end{equation}
To be sure I have brushed over important details, but
this equation conveys a great deal.  First, the overall
amplitude is set by the inverse of the growth factor,
which is just the ratio of the radiation energy density
to matter density at present.  Next, it explains why
theoretical cosmologists were so relieved when
the COBE DMR detected temperature fluctuations of this amplitude,
and conversely why one heard offhanded remarks before
the COBE DMR detection that the standard cosmology
was in trouble because the CBR temperature was too uniform
to allow for the observed structure to develop.
Finally, it illustrates one of the reasons
why cosmologists who study structure formation
have embraced the flat-Universe model with such
enthusiasm:  If we accept the Universe that meets the eye, $\Omega_0\sim
0.1$ and baryons only, then the simplest models of structure
formation predict temperature fluctuations of the order
of $10^{-3}$, far too large to be consistent with observation.
Later, I will mention Peebles' what-you-see-is-what-you-get
model \cite{pib}, also known as PIB for primeval baryon isocurvature
fluctuation, which is still viable because the spectrum of perturbations
decreases rapidly with scale so that the perturbations that
give rise to CBR fluctuations are small (which
is no mean feat).  Historically, it was
fortunate that one started with a low-$\Omega_0$, baryon-dominated
Universe:  the theoretical predictions for the CBR fluctuations
were sufficiently favorable that experimentalist
were stirred to try to measure them---and then, slowly,
theorists lowered their predictions.  Had the theoretical
expectations begun at $10^{-5}$, experimentalists might
have been too discouraged to even try!

\subsubsection{An initial data problem}
 
With the COBE DMR detection in hand we can praise
the success of the gravitational instability
scenario; however, the details now remain to be
filled in.  The structure formation problem is now one
of initial data, namely
\begin{enumerate}
\item The quantity and composition of
matter in the Universe, $\Omega_0$, $\Omega_B$, and $\Omega_{\rm other}$.

\item The spectrum of initial density perturbations:
for the purist, $(\delta\rho /\rho )_{\rm HOR}$, or
for the simulator, the Fourier amplitudes
at the epoch of matter-radiation equality.

\end{enumerate}
In a statistical sense, these initial data
provide the ``blueprint'' for the formation of structure.

The initial data are the challenge and the opportunity.  Although
the gravitational instability picture has been around since
the discovery of the CBR itself, the lack of specificity
in initial data has impeded progress.  With the advent
of the study of the earliest history of the Universe
a new door was opened.  We now have several well motivated
early-Universe blueprints:  Inflation-produced density
perturbations and nonbaryonic dark matter; cosmic-string
produced perturbations and nonbaryonic dark matter \cite{cs}; texture
produced density perturbations and nonbaryonic dark matter \cite{texture},
and one ``conventional model,''
a baryon-dominated Universe with isocurvature
fluctuations\footnote{Isocurvature baryon-number fluctuations correspond
at early times to fluctuations in the local baryon number
but not the energy density.  At late times, when the Universe
is matter dominated, they become fluctuations in the
mass density of a comparable amplitude.} \cite{pib}.
Structure formation provides the opportunity to probe the earliest
history of the Universe.  I will focus on
the cold dark matter ``family of models,'' which are
motivated by inflation.   Already the flood
of data has all but eliminated the conventional model; the
texture and cosmic-string models face severe problems with
CBR anisotropy---and who knows, even the cold dark matter
models may be eliminated.
 
\section{INFLATIONARY THEORY}

\subsection{Generalities}

As successful as the big-bang cosmology it suffers from a dilemma
involving initial data.  Extrapolating back, one finds that
the Universe apparently began from a very special state:
A slightly inhomogeneous and very flat Robertson-Walker spacetime.
Collins and Hawking showed that the
set of initial data that evolve to a spacetime that is
as smooth and flat as ours is today of measure zero \cite{collins}.
(In the context of simple grand unified theories, the hot big bang
suffers from another serious problem:  the extreme overproduction of superheavy
magnetic monopoles; in fact, it was an attempt to solve the
monopole problem which led Guth to inflation.)

The cosmological appeal of inflation is its ability to lessen the dependence
of the present state of the Universe upon the initial state.  Two
elements are essential to doing this:  (1) accelerated (``superluminal'')
expansion and the concomitant tremendous growth of the
scale factor; and (2) massive entropy production \cite{htw}.
Together, these two features allow a small, smooth subhorizon-sized
patch of the early Universe to grow to a large enough size and contain enough
heat (entropy in excess of $10^{88}$) to easily encompass our
present Hubble volume.  Provided that
the region was originally small compared to the curvature
radius of the Universe it would appear flat then and today (just
as any small portion of the surface of a sphere appears flat).

While there is presently no standard model of inflation---just as
there is no standard model for physics at these energies
(typically $10^{15}\GeV$ or so)---viable models have much in
common.  They are based upon well posed, albeit highly speculative,
microphysics involving the classical evolution of a scalar field.
The superluminal expansion is driven by the potential
energy (``vacuum energy'') that arises when the scalar field
is displaced from its potential-energy minimum, which results in
nearly exponential expansion.  Provided
the potential is flat, during the time it takes for the field to roll
to the minimum of its potential the Universe undergoes many e-foldings
of expansion (more than around 60 or so are required to realize
the beneficial features of inflation).
As the scalar field nears the minimum,  the vacuum energy has been
converted to coherent oscillations of the scalar field, which
correspond to nonrelativistic scalar-field particles.  The eventual
decay of these particles into lighter particles and their thermalization
results in the ``reheating'' of the Universe and accounts for all the
heat in the Universe today (the entropy production event).

Superluminal expansion and the tremendous growth of the scale
factor (by a factor greater than that since the end of inflation)
allow quantum fluctuations on very small scales ($\la 10^{-23}\rcm$)
to be stretched to astrophysical scales ($\ga 10^{25}\rcm$).
Quantum fluctuations in the scalar field responsible for inflation
ultimately lead to an almost scale-invariant spectrum
of density perturbations \cite{scalar}, and quantum fluctuations in
the metric itself lead to an almost scale-invariant spectrum of gravity-waves
\cite{tensor}.   Scale invariance for density perturbations means
scale-independent fluctuations in the gravitational potential
(equivalently, density perturbations of different wavelength
cross the horizon with the same amplitude);
scale invariance for gravity waves means that
gravity waves of all wavelengths cross the horizon with the same amplitude.
Because of subsequent evolution, neither the scalar nor the
tensor perturbations are scale invariant today.

\subsection{Metaphysical implications}

Inflation alleviates the ``specialness'' problem greatly, but does
not eliminate all dependence upon the initial state \cite{nohair}.
All open FRW models will inflate and become flat; however,
many closed FRW models will recollapse before they can inflate.
If one imagines the most general initial spacetime as being comprised
of negatively and positively curved FRW (or Bianchi) models that are
stitched together, the failure of the positively
curved regions to inflate is of little consequence:  because
of exponential expansion during inflation the negatively curved
regions will occupy most of the space today.
Nor does inflation solve the smoothness problem forever;
it just postpones the problem into the exponentially distant future:
We will be able to see outside our smooth inflationary
patch and $\Omega$ will
start to deviate significantly from unity at a time $t\sim t_0
\exp [3(N-N_{\rm min}]$, where $N$ is the actual number of e-foldings
of inflation and $N_{\rm min}\sim 60$ is the minimum required to
solve the horizon/flatness problems.

Linde has emphasized that inflation has changed our view
of the Universe in a very fundamental way \cite{eternal}.
While cosmologists have long used
the Copernican principle to argue that the Universe must be smooth
because of the smoothness of our Hubble volume, in the post-inflation
view, our Hubble volume is smooth because it is a small
part of a region that underwent inflation.  On
the largest scales the structure of the Universe is likely to be
very rich:  Different regions may have undergone different amounts of
inflation, may have different laws of physics because they
evolved into different vacuum states (of equivalent energy), and
may even have different numbers of spatial dimensions.  Since it is
likely that most of the volume of the Universe is still undergoing
inflation and that inflationary patches are being constantly produced
(eternal inflation), the age of the Universe is
a meaningless concept and our expansion age merely measures the
time back to the end of our inflationary event!

\subsection{Models}

In Guth's seminal paper \cite{guth} he introduced the idea of inflation,
sung its praises, and showed that the model that he based the idea
upon did not work!  Thanks to very important contributions by Linde
\cite{linde} and Albrecht and Steinhardt \cite{as} that was quickly
remedied, and today there are many viable models of inflation.
That of course is both good news and bad news; it means that there
is no standard model of inflation.  Again, the absence of a standard
model of inflation should be viewed in the light of our general
ignorance about fundamental physics at these energies.

Many different approaches have taken in constructing particle-physics
models for inflation.  Some have focussed on very simple scalar potentials,
e.g., $V(\phi ) = \lambda \phi^4$ or $=m^2\phi^2/2$, without regard
to connecting the model to any underlying theory \cite{chaotic,st}.
Others have proposed more complicated models that attempt to make
contact with speculations about physics at very high energies,
e.g., grand unification \cite{pi}, supersymmetry \cite{florida,olive,lbl},
preonic physics \cite{pati}, or supergravity \cite{liddle}.
Several authors have attempted
to link inflation with superstring theory \cite{banks} or ``generic
predictions'' of superstring theory such as pseudo-Nambu-Goldstone
boson fields \cite{unnatural}.  While the scale of the vacuum energy
that drives inflation is typically of order $(10^{15}\GeV)^4$, a
model of inflation at the electroweak scale, vacuum energy $\approx(1\TeV )^4$,
has been proposed \cite{knox}.  There are also models in which there are
multiple epochs of inflation \cite{multiple}.

In all of the models above gravity is described by
general relativity.  A qualitatively different approach is to
consider inflation in the context of alternative theories of
gravity.  (After all, inflation probably involves physics at
energy scales not too different from the Planck scale and the
effective theory of gravity at these energies could well be
very different from general relativity; in fact, there are some
indications from superstring theory that gravity in these
circumstances might be described by a Brans-Dicke like theory.)
Perhaps the most successful of these models is first-order inflation
\cite{lapjs,kolbreview}.  First-order inflation returns
to Guth's original idea of a strongly first-order
phase transition; in the context of general relativity Guth's model
failed because the phase transition, if inflationary, never completed.
In theories where the effective strength of gravity evolves, like
Brans-Dicke theory, the weakening of gravity during inflation
allows the transition to complete.  In other models based upon
nonstandard gravitation theory, the scalar field responsible for
inflation is itself related to the size of additional spatial dimensions,
and inflation then also explains why our three spatial dimensions are
so big, while the other spatial dimensions are so small.

All models of inflation have one feature in common:  the scalar
field responsible for inflation has a very flat potential-energy
curve and is very weakly coupled.  This typically leads to a very
small dimensionless number, usually a dimensionless coupling
of the order of $10^{-14}$.
Such a small number, like other small numbers in physics (e.g.,
the ratio of the weak to Planck scales $\approx 10^{-17}$ or
the ratio of the mass of the electron to the $W/Z$ boson masses $\approx
10^{-5}$), runs counter to one's belief that a truly fundamental
theory should have no tiny parameters, and cries out for an
explanation.  At the very least, this small number must be stabilized against
quantum corrections---which it is in all of the previously
mentioned models.\footnote{It is sometimes stated that inflation
is unnatural because of the small coupling of the scalar field
responsible for inflation; while the small coupling certainly begs
explanation, these inflationary models are not unnatural in
the rigorous technical sense as the small number is stable
against quantum fluctuations.}  In some models, the small number in the
inflationary potential is related to other small numbers in
particle physics:  for example, the ratio of the electron mass
to the weak scale or the ratio of the unification scale to
the Planck scale.  Explaining the origin of
the small number that seems to be associated with
inflation is both a challenge and an opportunity.

Because of the growing base of observations that bear on inflation,
another approach to model building is emerging:  the use of observations
to constrain the underlying inflationary potential.
I will return to ``reconstructing'' the inflationary
potential from data later.
Before going on, I want to emphasize that while there are many
varieties of inflation, there are robust predictions
which are crucial to sharply testing inflation.

\subsection{Three robust predictions}

Inflation makes three robust\footnote{Because theorists
are so clever, it is not possible nor prudent to use the word
immutable.  Models that violate any or all of these ``robust
predications'' can and have been constructed.} predictions:

\begin{enumerate}

\item {\bf Flat universe.}  Because solving the ``horizon''
problem (large-scale smoothness in spite of small particle
horizons at early times) and solving the ``flatness'' problem
(maintaining $\Omega$ very close to unity until the present epoch)
are linked geometrically \cite{eu,htw}, this is the most robust
prediction of inflation.  Said another way, it is the prediction
that most inflationists would be least willing to give up.
(Even so, models of inflation have been
constructed where the amount of inflation is tuned just to give
$\Omega_0$ less than one today \cite{pu}.)  Through the Friedmann
equation for the scale factor, flat implies that the total
energy density (matter, radiation, vacuum energy, ...) is
equal to the critical density.

\item {\bf Nearly scale-invariant spectrum of gaussian density perturbations.}
Essentially all inflation models predict a nearly, but
not exactly, scale-invariant
spectrum of gaussian density perturbations \cite{st}.  Described in terms
of a power spectrum, $P(k) \equiv \langle |\delta_k|^2 \rangle
= Ak^n$, where $\delta_k$ is the Fourier transform of the primeval
density perturbations, and the spectral index $n\approx 1$
(the scale-invariant limit is $n=1$).  The inflationary prediction
is statistical:  the $\delta_k$ are
drawn from a gaussian distribution whose variance is
$|\delta_k|^2$.  The overall amplitude $A$
is very model dependent.  Density perturbations give rise to CBR anisotropy
as well as seeding structure formation.
Requiring that the density perturbations are consistent
with the observed level of anisotropy of the CBR (and large enough
to produce the observed structure formation) is the most severe
constraint on inflationary models and leads to the small
dimensionless number that all inflationary models have.

\item {\bf Nearly scale-invariant spectrum of gravitational waves.}
These gravitational waves have wavelengths from ${\cal O}(1\km )$
to the size of the present Hubble radius and beyond.  Described in
terms of a power spectrum for the dimensionless gravity-wave amplitude
at early times, $P_T(k) \equiv \langle |h_k|^2 \rangle = A_Tk^{n_T-3}$, where
the spectral index $n_T \approx 0$ (the scale-invariant limit
is $n_T =0$).  As before, the power spectrum specifies
the variance of the Fourier components.
Once again, the overall amplitude $A_T$ is model dependent (varying
as the value of the inflationary vacuum energy).  Unlike density
perturbations, which are required to initiate structure formation,
there is no cosmological lower bound to the amplitude of
the gravity-wave perturbations.  Tensor perturbations
also give rise to CBR anisotropy; requiring that they do not lead to
excessive anisotropy implies that the energy density that drove
inflation must be less than about $(10^{16}\GeV )^4$.  This
indicates that if inflation took place, it did so at an energy well
below the Planck scale.\footnote{To be more precise, the part of inflation
that led to perturbations on scales within the present horizon involved
subPlanckian energy densities.  In
some models of inflation, the earliest stages, which do not influence
scales that we are privy to, involve energies as large as the Planck scale.}

\end{enumerate}

There are other interesting consequences of inflation that
are less generic.  For example, in models of first-order inflation,
in which reheating occurs through the nucleation and collision of
vacuum bubbles, there is an additional, larger amplitude, but
narrow-band, spectrum of gravitational waves ($\Omega_{\rm GW}h^2
\sim 10^{-6}$) \cite{vacuumpop}.  In other models large-scale primeval magnetic
fields of interesting size are seeded during inflation \cite{bfield}.

\section{Inflation:  The Details}

In this Section I discuss how to analyze an
inflationary model, given the scalar potential.
In two sections hence I will work through a
number of examples.  The focus will
be on the metric perturbations---density fluctuations \cite{scalar}
and gravity waves \cite{tensor}---that arise due to quantum fluctuations,
and the CBR temperature anisotropies that result
from them.\footnote{Isocurvature perturbations
can arise due to quantum fluctuations in other massless fields,
e.g., the axion field, if it exists \cite{isocurv}.}
Perturbations on all astrophysically interesting scales,
say $1\Mpc$ to $10^4\Mpc$, are produced during
an interval of about 8 e-folds around 50 e-folds
before the end of inflation, when these scales
crossed outside the horizon during inflation.
I will show how
the density perturbations and gravity waves
can be related to three features of the inflationary
potential:  its value $V_{50}$, its steepness
$x_{50}\equiv (\mpl V^\prime /V)_{50}$, and the change in
its steepness $x_{50}^\prime$, evaluated in
the region of the potential where the scalar
field was about 50 e-folds before the end of
inflation.  In principle, cosmological observations,
most importantly CBR anisotropy, can be used
to determine the characteristics of the
density perturbations and gravitational waves
and thereby $V_{50}$, $x_{50}$, and
$x_{50}^\prime$.
 
All viable models of inflation are of the
slow-rollover variety, or can be recast as 
such \cite{allslow}.  In slow-rollover
inflation a scalar field
that is initially displaced from the minimum 
of its potential rolls slowly to that minimum, and as it does
the cosmic-scale factor grows very rapidly.
Once the scalar field reaches the
minimum of the potential it oscillates about it, so that the large 
potential energy has been converted into coherent
scalar-field oscillations, corresponding to
a condensate of nonrelativistic scalar particles. 
The eventual decay of these particles into lighter 
particle states and their subsequent thermalization
lead to the reheating of the Universe to a
temperature $T_{\rm RH} \simeq \sqrt{\Gamma \mpl}$,
where $\Gamma$ is the decay width of the
scalar particle \cite{reheat,allslow}.
Here, I will focus on the classical evolution of
the inflaton field during the slow-roll phase and
the small quantum fluctuations in the inflaton field which
give rise to density perturbations and those
in the metric which give rise to gravity waves.

To begin, let us assume that the scalar field driving inflation is
minimally coupled so that its stress-energy tensor
takes the canonical form,
\begin{equation}
T_{\mu\nu} = \partial_\mu\phi \partial_\nu\phi
-{\cal L}g_{\mu \nu};
\end{equation}
where the Lagrangian density of the scalar field 
${\cal L} = {1\over 2}\partial_\mu\phi\partial^\mu\phi - V(\phi )$.
If we make the usual assumption that the scalar field
$\phi$ is spatially homogeneous, or at least so over a Hubble radius,
the stress-energy tensor takes the perfect-fluid form
with energy density, $\rho = {1\over 2}{\dot\phi}^2
+V(\phi )$, and isotropic pressure, $p={1\over 2}{\dot\phi}^2 -V(\phi )$.
The classical equations of motion
for $\phi$ can be obtained from the first law of thermodynamics,
$d(R^3\rho )=-pdR^3$, or by taking the four-divergence of $T^{\mu\nu}$:
\begin{equation}\label{eq:sr}
\ddot\phi + 3H\dot\phi + V^\prime (\phi ) = 0; 
\end{equation}
the $\Gamma \dot\phi$ term responsible for reheating has
been omitted since we shall only be interested in the
slow-rollover phase.  In addition, there is the
Friedmann equation, which governs the expansion of the Universe,
\begin{equation}\label{eq:feq}
H^2 = {8\pi \over 3\mpl^2}\left( V(\phi ) + {1\over 2} 
{\dot\phi}^2\right) \simeq {8\pi V(\phi )\over 3 \mpl^2}; 
\end{equation} 
where we assume that the contribution of 
all other forms of energy density, e.g., radiation and kinetic
energy of the scalar field, and
the curvature term ($k/R^2$) are negligible.
The justification for discussing inflation in the context
of a flat FRW model with a homogeneous scalar field
driving inflation were discussed earlier
(and at greater length in Ref.~\cite{inflation});
including the $\phi$ kinetic
term increases the righthand side of Eq. (\ref{eq:feq}) by a factor
of $(1+x^2/48\pi )$, a small correction for viable models.

In the next Section I will be more
precise about the amplitude of density perturbations and
gravitational waves; for now, let me briefly discuss how
these perturbations arise and give their characteristic amplitudes.
The metric perturbations produced in inflationary
models are very nearly ``scale invariant,''
a particularly simple spectrum which was first discussed
by Harrison and Zel'dovich \cite{hz}, and arise due
to quantum fluctuations.
In deSitter space all massless scalar fields experience
quantum fluctuations of amplitude $H/2\pi$.
The graviton is massless and can be described by
two massless scalar fields, $h_{+,\times} =
\sqrt{16\pi G}\phi_{+,\times}$ ($+$ and $\times$
are the two polarization states).  The inflaton by virtue of its flat
potential is for all practical purposes massless.

Fluctuations in the inflaton field lead to density fluctuations
because of its scalar potential, $\delta \rho \sim
HV^\prime$; as a given mode crosses outside the horizon, the density
perturbation on that scale becomes a classical metric
perturbation.  While outside the horizon, the description
of the evolution of a density perturbation is beset
with subtleties associated with the gauge freedom in
general relativity; there is, however, a simple gauge-invariant
quantity, $\zeta \simeq \delta\rho /(\rho +p)$, which
remains constant outside the horizon.  By equating the
value of $\zeta$ at postinflation horizon crossing with
its value as the scale crosses outside the horizon
it follows that $(\delta \rho /\rho )_{\rm HOR}
\sim HV^\prime /{\dot\phi}^2$ (note:  $\rho +p
= {\dot\phi}^2$); see Fig.~7.

The evolution of a gravity-wave perturbation is
even simpler; it obeys the massless Klein-Gordon
equation
\begin{equation}
{\ddot h}^i_k +3H{\dot h}^i_k + k^2h^i_k/R^2=0;
\end{equation}
where $k$ is the wavenumber of the mode and $i=+,\times$.  For
superhorizon sized modes, $k\la RH$, the solution
is simple:  $h^i_k=\,$const.  Like their density
perturbation counterparts, gravity-wave perturbations
become classical metric perturbations as they cross
outside the horizon; they are characterized by an
amplitude $h^i_k \simeq \sqrt{16\pi G}(H/2\pi )\sim
H/\mpl$.  At postinflation horizon crossing their
amplitude is unchanged.

Finally, let me write the horizon-crossing amplitudes
of the scalar and tensor metric perturbations
in terms of the inflationary potential,
\begin{eqnarray}\label{eq:scalartensor}
(\delta\rho /\rho)_{{\rm HOR},\lambda} & =  & c_S \left(
{V^{3/2}\over \mpl^3 V^\prime}\right)_1; \\
h_{{\rm HOR},\lambda} &  =  &
c_T \left({V^{1/2}\over \mpl^2}\right)_1 ;
\end{eqnarray}
where $(\delta\rho /\rho )_{{\rm HOR},\lambda}$ is the amplitude
of the density perturbation on the scale $\lambda$
when it crosses the Hubble radius during the post-inflation epoch,
$h_{{\rm HOR},\lambda}$ is the dimensionless amplitude
of the gravitational wave perturbation on the scale $\lambda$
when it crosses the Hubble radius, and $c_S$, $c_T$
are numerical constants of order unity.
Subscript 1 indicates that the quantity involving
the scalar potential is to be evaluated when the
scale in question crossed outside the
horizon during the inflationary era.  The metric
perturbations produced by inflation are characterized
by almost scale-invariant horizon-crossing amplitudes;
the slight deviations from scale invariance result
from the variation of $V$ and $V^\prime$ during inflation
which enter through the dependence upon $t_1$.
[In Eq. (\ref{eq:scalartensor})
I got ahead of myself and used the slow-roll
approximation (see below)
to rewrite the expression, $(\delta \rho /\rho )_{{\rm HOR},
\lambda}\simeq HV^\prime /\dot\phi$, in terms of the potential only.]

Eqs. (\ref{eq:sr}-\ref{eq:scalartensor}) are the
fundamental equations that govern
inflation and the production of metric perturbations.
It proves very useful to recast these equations using
the scalar field as the independent variable; we
then express the scalar and tensor perturbations
in terms of the value of the potential, its steepness,
and the rate of change of its steepness when the interesting
scales crossed outside the Hubble radius during inflation,
about 50 e-folds in scale factor before the end of inflation,
defined by
$$V_{50} \equiv V(\phi_{50}); \qquad
x_{50} \equiv {\mpl V^\prime (\phi_{50})\over
V(\phi_{50})}; \qquad x_{50}^\prime = {\mpl V^{\prime\prime}
(\phi_{50})\over V(\phi_{50})} - {\mpl [V^\prime (\phi_{50})]^2
\over V^2(\phi_{50})}.$$

To evaluate these three quantities 50 e-folds before
the end of inflation we must find the value of
the scalar field at this time.
During the inflationary phase the $\ddot\phi$ term is
negligible (the motion of $\phi$ is friction dominated), 
and Eq. (\ref{eq:sr}) becomes
\begin{equation}\label{eq:sra}
\dot\phi \simeq {-V^\prime (\phi) \over 3H}; 
\end{equation} 
this is known as the slow-roll approximation \cite{st}.
While the slow-roll approximation is almost
universally applicable, there are models where
the slow-roll approximation cannot be used;
e.g., a potential where during the crucial 8 e-folds
the scalar field rolls uphill, ``powered'' by the
velocity it had when it hit the incline.

The conditions that must be satisfied in order that
$\ddot\phi$ be negligible are:
\begin{eqnarray}\label{eq:condx}
|V^{\prime\prime}| & < & 9H^2 \simeq 24\pi V/\mpl^2; \\
|x| \equiv |V^\prime \mpl /V|  & <  & \sqrt{48\pi}.
\end{eqnarray}
The end of the slow roll occurs 
when either or both of these inequalities 
are saturated, at a value of $\phi$ denoted by $\phi_{\rm end}$.
Since $H\equiv {\dot R} /R$,
or $Hdt = d\ln R$, it follows that
\begin{equation} 
d\ln R = {8\pi\over \mpl^2}\, {V(\phi ) d\phi 
\over -V^\prime (\phi )} = -{8\pi d\phi \over \mpl \,x}. 
\end{equation} 
Now express the cosmic-scale factor
in terms of is value at the end of inflation, $R_{\rm end}$,
and the number of e-foldings before the end of inflation, $N(\phi )$,
$$R = \exp [-N(\phi )]\,R_{\rm end}. $$
The quantity $N(\phi )$ is a time-like variable whose
value at the end of inflation is zero and whose
evolution is governed by
\begin{equation}\label{eq:neq}
{dN \over d\phi} = {8\pi \over \mpl\,x} .
\end{equation}
Using Eq. (\ref{eq:neq}) we can compute the value of
the scalar field 50 e-folds before the end of inflation ($\equiv
\phi_{50}$); the values of
$V_{50}$, $x_{50}$, and $x_{50}^\prime$ follow directly.
 
As $\phi$ rolls down its potential during inflation its
energy density decreases,
and so the growth in the scale factor is not exponential.
By using the fact that the stress-energy of the scalar 
field takes the perfect-fluid form, we can solve
for evolution of the cosmic-scale factor.  Recall,
for the equation of state $p=\gamma \rho$, the scale factor grows as 
$R\propto t^q$, where $q= 2/3(1+\gamma )$.  Here,
\begin{eqnarray}\label{eq:eos}
\gamma & = & {{1\over 2}{\dot\phi}^2 - V \over
        {1\over 2}{\dot\phi}^2 +V} = { x^2-48\pi \over 
        x^2 +48\pi };  \\
q & = & {1\over 3} + {16\pi \over x^2}.
\end{eqnarray}
Since the steepness of the potential can change during
inflation, $\gamma$ is not in general constant;
the power-law index $q$ is more
precisely the logarithmic rate of the change of the
logarithm of the scale factor, $q=d\ln R/d\ln t$.

When the steepness parameter is small,
corresponding to a very flat potential, $\gamma$ is close 
to $-1$ and the scale factor grows as a very large 
power of time.  To solve the horizon problem the 
expansion must be ``superluminal'' (${\ddot R}>0$), 
corresponding to $q>1$, which requires that $x^2< 24\pi$.
Since ${1\over 2}{\dot\phi}^2/V = x^2/48\pi$,
this implies that ${1\over 2}{\dot\phi}^2 /V(\phi ) < {1\over 2}$,
justifying neglect of the scalar-field kinetic energy
in computing the expansion rate for all but the
steepest potentials.  (In fact there are much stronger
constraints; the COBE DMR data imply that $n \ga 0.5$,
which restricts $x_{50}^2 \la 4\pi$, ${1\over 2}{\dot\phi}^2/V
\la {1\over 12}$, and $q\ga 4$.)
 
Next, let us relate the size of a given scale to
when that scale crosses outside
the Hubble radius during inflation, specified by $N_1(\lambda )$,
the number of e-folds before the end of inflation.
The physical size of a perturbation is related to its comoving size,
$\lambda_{\rm phys}=R\lambda$; with the usual
convention, $R_{\rm today} =1$, the comoving size
is the physical size today.  When the scale $\lambda$
crosses outside the Hubble radius $R_1
\lambda = H_1^{-1}$.  We then assume that:  (1) at the end of
inflation the energy density is ${\cal M}^4\simeq
V(\phi_{\rm end})$; (2) inflation is followed by a
period where the energy density of the Universe is dominated by coherent
scalar-field oscillations which decrease as
$R^{-3}$; and (3) when value of the scale factor
is $R_{\rm RH}$ the Universe reheats to a temperature
$T_{\rm RH} \simeq \sqrt{\mpl \Gamma}$ and expands
adiabatically thereafter.  The ``matching equation''
that relates $\lambda$ and $N_1(\lambda )$ is:
\begin{equation}\label{eq:prematch}
\lambda =  {R_{\rm today}\over R_1 } H_1^{-1}
        = {R_{\rm today} \over R_{\rm RH}} \,
        {R_{\rm RH}\over R_{\rm end}}\,
        {R_{\rm end}\over R_1}\, H_1^{-1}.
\end{equation} 
Adiabatic expansion since reheating implies
$R_{\rm today}/R_{\rm RH} \simeq T_{\rm RH}/2.73\,{\rm K}$;
and the decay of the coherent scalar-field oscillations
implies $(R_{\rm RH}/R_{\rm end})^3 = ({\cal M}/T_{\rm RH})^4$.
If we define ${\bar q} = \ln (R_{\rm end}/R_1)/\ln (t_{\rm end}/t_1)$,
the mean power-law index, it follows that $(R_{\rm end}/R_1)H_1^{-1} =
\exp [N_1({\bar q}-1)/{\bar q}]H_{\rm end}^{-1}$,
and Eq. (\ref{eq:prematch}) becomes
\begin{equation}\label{eq:match}
N_1(\lambda )=  {{\bar q}\over {\bar q} -1}\,
\left[48+\ln \lambda_{\Mpc}
+ {2\over 3}\ln({\cal M}/10^{14}\GeV) + {1\over 3}
\ln (T_{\rm RH}/10^{14}\GeV ) \right];
\end{equation}
In the case of perfect reheating, which probably only applies to
first-order inflation, $T_{\rm RH} \simeq{\cal M}$.

The scales of astrophysical interest
today range roughly from that of galaxy size,
$\lambda \sim \Mpc$, to the present
Hubble scale, $H_0^{-1}\sim 10^4\Mpc$; up to the
logarithmic corrections these
scales crossed outside the horizon between about 
$N_1(\lambda )\sim 48$ and $N_1(\lambda )\simeq 56$
e-folds before the end of inflation.  {\it That is,
the interval of inflation that determines its all observable
consequences covers only about 8 e-folds.}

Except in the case of strict power-law inflation
$q$ varies during inflation; this means that the
$(R_{\rm end}/R_1)H_1^{-1}$ factor in Eq. (\ref{eq:prematch})
cannot be written in closed form.
Taking account of this, the matching equation becomes
a differential equation,
\begin{equation}\label{eq:dmatch}
{d\ln\lambda_{\Mpc}\over dN_1} = {q(N_1) -1  \over q(N_1)};
\end{equation}
subject to the ``boundary condition:''
$$\ln \lambda_{\Mpc}
= -48 -{4\over 3}\ln ({\cal M}/10^{14}\GeV )+{1\over 3}\ln
(T_{\rm RH}/10^{14}\GeV )$$
for $N_1=0$, the matching relation
for the mode that crossed outside the Hubble radius at the
end of inflation.  Equation (\ref{eq:dmatch}) allows
one to obtain the precise expression for when a given scale
crossed outside the Hubble radius during inflation.  To
actually solve this equation, one would need to supplement it
with the expressions $dN/d\phi = 8\pi /\mpl x$ and
$q = 16\pi /x^2$.  For our purposes we need only know:  (1) The scales of
astrophysical interest correspond to $N_1\sim ``50\pm 4$,''
where for definiteness we will throughout take this
to be an equality sign.  (2) The expansion of Eq.
(\ref{eq:dmatch}) about $N_1 =50$,
\begin{equation}\label{eq:dndl}
\Delta N_1(\lambda ) = \left( {q_{50}-1\over q_{50} } \right)
\Delta \ln \lambda_{\Mpc} ;
\end{equation}
which, with the aid of Eq. (\ref{eq:neq}), implies that
\begin{equation} \label{eq:philambda}
\Delta \phi = \left( {q_{50} -1\over q_{50}}\right)\,
{x_{50}\over 8\pi}\, \Delta \lambda_{\Mpc} .
\end{equation}

We are now ready to express the perturbations
in terms of $V_{50}$, $x_{50}$, and
$x_{50}^\prime$.  First, we must solve for the value of $\phi$,
50 e-folds before the end of inflation.
To do so we use Eq. (\ref{eq:neq}),
\begin{equation}\label{eq:phi50}
N(\phi_{50} ) = 50 = {8\pi \over \mpl^2}\int_{\phi_{\rm end}}^{\phi_{50}} 
{Vd\phi \over V^\prime}. 
\end{equation} 
Next, with the help of Eq. (\ref{eq:philambda})
we expand the potential $V$ and its steepness
$x$ about $\phi_{50}$:
\begin{equation}\label{eq:expandV}
V \simeq V_{50} + V_{50}^\prime (\phi - \phi_{50} )
= V_{50}\left[ 1 +  {x_{50}^2\over 8\pi}\,\left( {q_{50}\over q_{50}-1}
\right) \,  \Delta\ln \lambda_{\Mpc} \right] ;
\end{equation} 
\begin{equation}\label{eq:expandx}
x \simeq x_{50} + x^\prime_{50}(\phi - \phi_{50})
= x_{50}\left[ 1 + {\mpl x_{50}^{\prime}
\over 8\pi}\,\left( {q_{50} \over q_{50}-1}\right)
\,\Delta\ln\lambda_{\Mpc} \right];
\end{equation}
of course these expansions only make sense for
potentials that are smooth.  We note that additional
terms in either expansion are ${\cal O}(\alpha_i^2)$ and
beyond the accuracy we are seeking.

Now recall the equations for the amplitude of the
scalar and tensor perturbations,
\begin{eqnarray}\label{eq:perts}
(\delta \rho /\rho )_{{\rm HOR},\lambda} & = & c_S \left( {V^{1/2}\over
\mpl^2 x}\right)_1 ;\\
h_{{\rm HOR},\lambda} & = & c_T\left( {V^{1/2}\over \mpl^2} \right)_1 ;
\end{eqnarray}
where subscript 1 means that the quantities are to be
evaluated where the scale $\lambda$ crossed outside the
Hubble radius, $N_1(\lambda )$ e-folds before the
end of inflation.   The origin of any deviation from
scale invariance is clear:  For tensor perturbations it
arises due to the variation of the potential; and for
scalar perturbations it arises due to the
variation of both the potential and its steepness.

Using Eqs. (\ref{eq:dndl}-\ref{eq:perts}) it is now
simple to calculate the power-law exponents $\alpha_S$
and $\alpha_T$ that quantify the deviations from scale
invariance,
\begin{eqnarray}\label{eq:alpha}
\alpha_T  &  =  & {x_{50}^2\over 16\pi}\,{q_{50}\over q_{50}-1} \simeq
        {x_{50}^2 \over 16\pi};  \\
\alpha_S &  =  &  \alpha_T - {\mpl x_{50}^\prime \over 8\pi}\,{q_{50}
\over q_{50} -1} \simeq {x_{50}^2\over 16\pi}
- {\mpl x_{50}^\prime \over 8\pi};
\end{eqnarray}
where
\begin{eqnarray}\label{eq:defofsi}
q_{50} &  =  &  {1\over 3} + {16\pi \over x_{50}^2}
        \simeq {16\pi \over x_{50}^2} ;\\
h_{{\rm HOR},\lambda} & = & c_T \left( {V_{50}^{1/2}\over
\mpl^2 } \right) \, \lambda_{\Mpc}^{\ \ \,\alpha_T};\\
(\delta\rho /\rho )_{{\rm HOR},\lambda} &  =  &  c_S
\left({V_{50}^{1/2}\over x_{50}\mpl^2}\right)\,
\lambda_{\Mpc}^{\ \ \,\alpha_S}.
\end{eqnarray}
The spectral indices $\alpha_i$ are defined as,
$\alpha_S = [d\ln (\delta\rho /\rho)_{{\rm HOR},
\lambda} / d\ln \lambda_{\Mpc}]_{50}$ and $\alpha_T =
[d\ln h_{{\rm HOR},\lambda} /d\ln
\lambda_{\Mpc}]_{50}$, and in general vary
slowly with scale.  Note too that the
deviations from scale invariance, quantified
by $\alpha_S$ and $\alpha_T$, are of the order of
$x_{50}^2$, $\mpl x_{50}^\prime$.  In the
expressions above we retained only lowest-order
terms in ${\cal O}(\alpha_i)$.  The next-order contributions to the
spectral indices are ${\cal O}(\alpha_i^2)$; those
to the amplitudes are ${\cal O}(\alpha_i)$ and
are given two sections hence.  The justification for
truncating the expansion at lowest order is that
the deviations from scale invariance are expected to be small---and
are required by astrophysically data to be small.

As I discuss in more detail two sections hence, the more
intuitive power-law indices $\alpha_S$, $\alpha_T$ are
related to the indices that are usually used
to describe the power spectra of scalar and tensor
perturbations, $P_S(k) \equiv |\delta_k|^2 = A k^n$
and $P_T(k) \equiv |h_k|^2 = A_Tk^{n_T}$,
\begin{eqnarray}\label{eq:indices}
n & = & 1-2\alpha_S = 1 -{x_{50}^2\over 8\pi}
+ {\mpl x_{50}^\prime\over 4\pi} ; \\
n_T & = & -2\alpha_T =  -{x_{50}^2\over 8\pi} .\\
\end{eqnarray}

CBR temperature fluctuations on large-angular scales
($\theta \ga 1^\circ$) due to metric perturbations
arise through the Sachs-Wolfe effect; very roughly, the
temperature fluctuation on a given angular scale
$\theta$ is related to the metric fluctuation on the
length scale that subtends that angle at last scattering,
$\lambda \sim 100h^{-1}\Mpc (\theta /{\rm deg})$,
\begin{eqnarray}
\left({\delta T\over T}\right)_\theta & \sim &
\left({\delta \rho \over \rho}\right)_{{\rm HOR},\lambda};\\
\left({\delta T\over T}\right)_\theta & \sim &
h_{{\rm HOR},\lambda} ;
\end{eqnarray}
where the scalar and tensor contributions to the CBR
temperature anisotropy on a given scale add in quadrature.
Let me be more specific about the amplitude of the
quadrupole CBR anisotropy.   For small $\alpha_S$,
$\alpha_T$ the contributions of each to
the quadrupole CBR temperature anisotropy:
\begin{eqnarray}\label{eq:quadanisotropy}
\left({\Delta T\over T_0}\right)_{Q-S}^2  &  \approx  &
{32\pi\over 45}{V_{50}\over \mpl^4 x_{50}^2};\\
\left({\Delta T\over T_0}\right)_{Q-T}^2  & \approx  &
0.61{V_{50}\over \mpl^4};
\end{eqnarray}
\begin{equation}\label{eq:ratio}
{T\over S}  \equiv {(\Delta T/T_0)_{Q-T}^2 \over
(\Delta T/T_0)_{Q-S}^2} \approx {0.28  x_{50}^2};
\end{equation}
where expressions have been evaluated
to lowest order in $x_{50}^2$ and $\mpl x_{50}^\prime$.
In terms of the spherical-harmonic expansion of the CBR
temperature anisotropy, the square of the quadrupole anisotropy is
defined to be $\sum_{m=-2}^{m=2}|a_{2m}|^2/4\pi$.

So what are these quantities precisely?  Inflation makes statistical
predictions.  The underlying density perturbations are
gaussian and the expression for $|\delta_k|^2$ is simply
the variance of the gaussian distribution for $\delta_k$.
Because the predicted multipole amplitudes $a_{lm}$ depend linearly
upon $\delta_k$ and $h_k$, the distribution of multipole amplitudes
is gaussian, with variance $\equiv \langle |a_{lm}|^2\rangle$.  This
underlying variance is comprised of scalar and tensor contributions.

How accurately can one hope to estimate the actual variance
of the underlying distribution?  If one had an ensemble of
observers distributed throughout the Universe who each measured
the CBR anisotropy at their position, then one could determine
the underlying variance to arbitrary precision by averaging their
$|a_{lm}|^2$'s  (hence the notation $\langle |a_{lm}|^2\rangle$
for the underlying variance).  However, we are privy to but
one CBR sky and for multipole $l$, only $2l+1$ multipole amplitudes.
Thus, we can only estimate the actual variance with finite precision.
This is nothing other than ordinary sampling
variance, but is often called ``cosmic variance.'' The sampling
variance of $\langle |a_{lm}|^2\rangle$---which is the irreducible
uncertainty in measuring $\langle |a_{lm}|^2\rangle$---is
simply given by $2\langle |a_{lm}|^2\rangle^2/(2l+1)$. The distribution
of the measured value of $\langle |a_{lm}|^2\rangle_{\rm MEAS}$
is just the $\chi^2$ distribution for $2l+1$ degrees of freedom.

Before going on, some general
remarks \cite{turner}.  The steepness parameter $x_{50}^2$ must
be less than about $24\pi$ to ensure
superluminal expansion.   For ``steep'' potentials,
the expansion rate is ``slow,'' i.e., $q_{50}$ closer to unity,
the gravity-wave contribution to the quadrupole CBR temperature anisotropy
becomes comparable to, or greater than, that of density
perturbations, and both scalar and tensor
perturbations exhibit significant deviations from scale invariance.
For ``flat'' potentials, i.e., small $x_{50}$,
the expansion rate is ``fast,'' i.e., $q_{50} \gg 1$,
the gravity-wave contribution to the quadrupole CBR temperature anisotropy
is much smaller than that of density perturbations, and
the tensor perturbations are scale invariant.  Unless
the steepness of the potential changes rapidly, i.e.,
large $x_{50}^\prime$, the scalar perturbations are also scale invariant.

\subsection{Metric perturbations and CBR anisotropy}

I was purposefully vague when discussing
the amplitudes of the scalar and tensor modes, except
when specifying their contributions to the quadrupole CBR temperature
anisotropy; in fact, the spectral indices $\alpha_S$ and
$\alpha_T$, together with the scalar and tensor
contributions to the CBR quadrupole serve to
provide all the information necessary.  Here I will
fill in more details about the metric perturbations.

The scalar and tensor metric perturbations are expanded
in harmonic functions, in the flat Universe predicted
by inflation, plane waves,
\begin{eqnarray}
h_{\mu\nu}({\bf x}, t) & = & {1\over (2\pi )^3}
\int d^3k\,h_{\bf k}^i (t)\, \varepsilon_{\mu\nu}^i
\, e^{-i{\bf k}\cdot{\bf x}} ;\\
{\delta\rho ({\bf x},t) \over \rho} & = &
{1\over (2\pi )^3}\int d^3k\,\delta_{\bf k} (t) \, e^{-i{\bf k}\cdot{\bf x}} ;
\end{eqnarray}
where $h_{\mu\nu}=R^{-2}g_{\mu\nu} - \eta_{\mu\nu}$,
$\varepsilon_{\mu\nu}^i$ is the polarization tensor
for the gravity-wave modes, and $i= +$, $\times$ are
the two polarization states.  Everything of interest
can be computed in terms of $h_{\bf k}^i$ and $\delta_{\bf k}$.
For example, the {\it rms} mass fluctuation
in a sphere of radius $r$ is obtained in terms of the
window function for a sphere and the power spectrum $P_S(k)
\equiv \langle |\delta_{\bf k}|^2\rangle$ (see below),
\begin{equation}
\langle (\delta M /M)^2\rangle_r = {9\over 2\pi^2r^2}\,
\int_0^\infty [j_1(kr)]^2 \,P_S(k) dk ;
\end{equation}
where $j_1(x)$ is the spherical Bessel function of first order.
If $P_S(k)$ is a power law, it follows roughly that
$(\delta M/M)^2 \sim k^3|\delta_{\bf k}|^2$,
evaluated on the scale $k=r^{-1}$.
This is what I meant by $(\delta \rho /
\rho)_{{\rm HOR},\lambda}$:  the {\it rms} mass
fluctuation on the scale $\lambda$
when it crossed inside the horizon.  Likewise,
by $h_{{\rm HOR},\lambda}$ I meant the {\it rms} strain
on the scale $\lambda$ as it crossed inside the Hubble radius,
$(h_{{\rm HOR},\lambda})^2 \sim k^3|h_{\bf k}^i|^2$.

In the previous discussions I have chosen to specify
the metric perturbations for the different Fourier
modes when they crossed inside the horizon,
rather than at a common time.  I did so because
scale invariance is made manifest, as the scale independence
of the metric perturbations at post-inflation horizon crossing.
Recall, in the case of scalar perturbations
$(\delta \rho /\rho )_{\rm HOR}$ is up to a numerical factor
the fluctuation in the Newtonian potential, and, by specifying
the scalar perturbations at horizon crossing, we avoid the
discussion of scalar perturbations on superhorizon
scales, which is beset by the subtleties associated with
the gauge noninvariance of $\delta_{\bf k}$.

It is, however, necessary to specify the perturbations at a common time
to carry out most calculations; e.g., an $N$-body simulation
of structure formation or the calculation of CBR anisotropy.
To do so, one has to
take account of the evolution of the perturbations
after they enter the horizon.
After entering the horizon tensor perturbations behave like
gravitons, with $h_{\bf k}$ decreasing as $R^{-1}$ and
the energy density associated with a given mode, $\rho_k \sim
\mpl^2 k^5|h_{\bf k}|^2/R^2$, decreasing as $R^{-4}$.  The evolution
of scalar perturbations is slightly more complicated; modes that
enter the horizon while the Universe is still radiation dominated
remain essentially constant until the Universe becomes matter
dominated (growing only logarithmically);
modes that enter the horizon after the Universe becomes
matter dominated grow as the scale factor.
(The gauge noninvariance of $\delta_{\bf k}$ is not an important
issue for subhorizon size modes.  A Newtonian analysis
suffices, and there is only one growing mode, corresponding to
a density perturbation.)

The method for characterizing the scalar perturbations
is by now standard:  The spectrum of
perturbations is specified at the present
epoch (assuming linear growth for all scales); the spectrum at earlier
epochs can be obtained by multiplying $\delta_{\bf k}$
by $R(t)/R_{\rm today}$.  The inflationary metric perturbations
are gaussian;
thus $\delta_{\bf k}$ is a gaussian, random variable.  Its
statistical expectation value is
\begin{equation}
\langle \delta_{\bf k}\,\delta_{\bf q}\rangle
= P_S(k)(2\pi )^3 \delta^{(3)} ({\bf k} -{\bf q});
\end{equation}
where the power spectrum today is written as
\begin{equation}
P_S(k) \equiv Ak^n T(k)^2;
\end{equation}
$n=1-2\alpha_S$ ($=1$ for scale-invariant perturbations), and
$T(k)$ is the ``transfer function'' which encodes
the information about the post-horizon crossing evolution
of each mode and depends
upon the matter content of the Universe, e.g., baryons plus cold dark
matter, baryons plus hot dark matter, baryons plus
hot and cold dark matter, and so on.
The transfer function is defined so that $T(k)\rightarrow 1$
for $k\rightarrow 0$ (long-wavelength perturbations); an
analytic approximation to the cold dark matter transfer
function is given by \cite{stat}
\begin{equation}\label{eq:cdmtf}
T(k)   =  {\ln (1+2.34 q)/2.34q \over \left[ 1 + (3.89q) +(16.1q)^2
+ (5.46q)^3 + (6.71q)^4 \right]^{1/4}} ;
\end{equation}
where $q= k/(\Omega_0 h^2 \Mpc^{-1})$.  Inflationary
power spectra for different dark matter possibilities
are shown in Fig.~9.

The overall normalization factor
\begin{equation}\label{eq:scalarnorm}
A = {1024\pi^3 \over 75H_0^{4}}\,{V_{50} \over \mpl^4 x_{50}^2}
\,{[1+{7\over 6}n_T - {1\over 3}(n-1)]
\left\{ \Gamma [{3\over 2} -{1\over 2}(n-1)]\right\}^2
\over 2^{n-1} [\Gamma ({3\over 2})]^2}\,k_{50}^{1-n} ;
\end{equation}
where the ${\cal O}(\alpha_i)$ correction to $A$
has been included \cite{ls}.  The quantity $n_T =-2\alpha_T =
-x_{50}^2/8\pi$, $n-1 = -2\alpha_S = n_T +x_{50}^\prime /4\pi$,
$k_{50}$ is the comoving wavenumber of the scale
that crossed outside the horizon 50 e-folds before the
end of inflation.  All the formulas below simplify if this
scale corresponds to the present horizon scale,
specifically, $k_{50}=H_0/2$.  [Eq. (\ref{eq:scalarnorm})
can be simplified by expanding $\Gamma ({3\over 2} + x )
= \Gamma (3/2)[1+x
(2-2\ln 2 -\gamma)]$, valid for $|x|\ll 1$;
$\gamma \simeq 0.577$ is Euler's constant.]

\begin{figure}
\vspace{4.6in}
\caption[power]
{Comparison of the cold dark matter perturbation spectrum
with CBR anisotropy measurements (boxes) and the distribution
of galaxies today (triangles).  Wavenumber $k$ is
related to length scale, $k=2\pi /\lambda$; error flags
are not shown for the galaxy distribution.  The
curve labeled MDM is hot + cold dark matter (``5\,eV''
worth of neutrinos); the other two curves are cold dark
matter models with Hubble constants of $50\kms\Mpc$ (labeled
CDM) and $35\kms\Mpc$.  (Figure courtesy of M.~White.)}
\end{figure}

From this expression it is simple to compute the
Sachs-Wolfe contribution of scalar perturbations
to the CBR temperature anisotropy; on angular scales much greater
than about $1^\circ$ (corresponding to multipoles $l \ll 100$) it
is the dominant contribution.  It is useful to expand the
CBR temperature on the sky in spherical harmonics,
\begin{equation}
{\delta T(\theta ,\phi )\over T_0} = \sum_{l\ge 2, m=-l}^{l=\infty ,m=l}
a_{lm}Y_{lm}(\theta ,\phi );
\end{equation}
where $T_0=2.73\,$K is the CBR temperature today and the dipole
term is subtracted out because it is cannot be separated from that
arising due to our motion with respect to the cosmic rest frame.
The predicted variance due to scalar perturbations is given by
\begin{eqnarray}\label{eq:scalarlm}
\langle |a_{lm}|^2\rangle & = & {H_0^4\over 2\pi}\,
\int_0^\infty k^{-2}\,P_S(k)\,|j_l(k r_0)|^2\,dk ; \\
&  \simeq &  {A2^{n-1}H_0^{4}\, r_0^{1-n}\over 16}\,
{\Gamma (l+{1\over 2}n-{1\over 2})
\Gamma (3-n) \over \Gamma (l-{1\over 2}n +{5\over 2})
[\Gamma (2-{1\over 2}n)]^2} ;
\end{eqnarray}
where $r_0\approx 2H_0^{-1}$ is the comoving distance to the last scattering
surface, and this expression is for the Sachs-Wolfe contribution from scalar
perturbations only.  For $n$ not too different from
one the expectation for the square of the quadrupole anisotropy is
\begin{equation}
\left( {\Delta T\over T_0} \right)_{Q-S}^2 \equiv
{5 |a_{2m}|^2 \over 4\pi } \approx {32\pi\over 45}\,
{V_{50} \over \mpl^4\,x_{50}^2}\,(k_{50}r_0)^{1-n}.
\end{equation}
(By choosing $k_{50}= r_0^{-1}= {1\over 2}H_0$,
the last factor becomes unity.)

The ensemble expectation for the multipole amplitudes
is often referred to as the angular power spectrum because
they encode the full information about predicted CBR anisotropy.
For example, the {\it rms} temperature fluctuation on a
given angular scale is related to the multipole amplitudes
\begin{equation}
\left( {\Delta T\over T}\right)^2_\theta \sim
l^2\langle |a_{lm}|^2\rangle \qquad {\rm for}\ \ l \simeq 200^\circ /\theta .
\end{equation}

The procedure for specifying the tensor modes is similar,
cf. Refs.~\cite{aw,white}.  For the modes that enter the
horizon after the Universe becomes matter dominated,
$k\la 0.1h^2\Mpc$, which are the only modes that contribute
significantly to CBR anisotropy on angular scales
greater than a degree,
\begin{equation}
h_{\bf k}^i (\tau ) = a^i ({\bf k}) \left( { 3j_1(k\tau )\over
k\tau }  \right) ;
\end{equation}
where $\tau = r_0(t/t_0)^{1/3}$ is conformal time.
[For the modes that enter the horizon during the radiation-dominated
era, $k \ga 0.1h^2\Mpc^{-1}$, the factor
$3j_1(k\tau )/k\tau$ is replaced by $j_0(k\tau )$
for the remainder of the radiation era.
In either case, the factor involving the spherical Bessel
function quantifies the fact that tensor perturbations
remain constant while outside the horizon, and after
horizon crossing decrease as $R^{-1}$.]

The tensor perturbations too are characterized by
a gaussian, random variable, here written as $a^i({\bf k})$;
the statistical expectation
\begin{equation}
\langle h_{\bf k}^i h_{\bf q}^j \rangle =
P_T (k)(2\pi )^6 \delta^{(3)} ({\bf k} -{\bf q})\delta_{ij};
\end{equation}
where the power spectrum
\begin{eqnarray}
P_T(k) & = & A_T k^{n_T -3} \left[ {3j_1(k\tau )\over k\tau}\right]^2 ; \\
A_T & = & {8 \over 3\pi }\, {V_{50} \over \mpl^4}\,
{(1+{5\over 6}n_T)[\Gamma ({3\over 2} - {1\over 2}n_T)]^2 \over
2^{n_T} [\Gamma ({3\over 2})]^2}\, k_{50}^{-n_T} ;
\end{eqnarray}
where the ${\cal O}(\alpha_i)$ correction to $A_T$ has been
included.  Note that $n_T = -2\alpha_T$ is zero
for scale-invariant perturbations.

Finally, the contribution of tensor perturbations to
the multipole amplitudes, which arise solely due
to the Sachs-Wolfe effect \cite{SW,aw,white}, is given by
\begin{equation}\label{eq:tensorlm}
\langle |a_{lm}|^2 \rangle \simeq 36 \pi^2 \,{\Gamma (l+3)
\over \Gamma (l-1) }\, \int_0^\infty\,
k^{n_T+1}\,A_T \, |F_l(k)|^2\,dk ;
\end{equation}
where
\begin{equation}
F_l(k)  =  - \int_{r_D}^{r_0} \, dr \,
{j_2(kr)\over kr}\,\left[ {j_l(kr_0-kr) \over
(kr_0-kr)^2}\right] ;
\end{equation}
and $r_D = r_0/(1+z_D)^{1/2} \approx r_0/35$ is
the comoving distance to the horizon at
decoupling (= conformal time at decoupling).
Equation (\ref{eq:tensorlm}) is approximate
in that very short wavelength modes, $kr_0\gg 100$,
that crossed inside the horizon before matter-radiation
equality have not been properly taken into account; to
take them into account, the integrand must be multiplied
by a transfer function,
\begin{equation}
T(k) \simeq 1.0 + 1.44(k/k_{\rm EQ}) +2.54 (k/k_{\rm EQ})^2;
\end{equation}
where $k_{\rm EQ} \equiv H_0/(2\sqrt{2}-2)R_{\rm EQ}^{1/2}$
is the scale that entered the horizon at matter radiation
equality \cite{turner}.  In addition, for $l\ga 1000$, the
finite thickness of the last-scattering surface
must be taken into account.

The tensor contribution to the quadrupole CBR temperature
anisotropy for $n_T$ not too different from zero is
\begin{equation}
\left( {\Delta T\over T_0} \right)_{Q-T}^2 \equiv
{5|a_{2m}|^2\over 4\pi} \simeq 0.61 {V_{50}\over \mpl^4}\, (k_{50}r_0)^{-n_T};
\end{equation}
where the integrals in the previous expressions have been evaluated numerically.

Both the scalar and tensor contributions to a given
multipole are dominated by wavenumbers $kr_0\sim l$.
For scale-invariant perturbations and small $l$,
both the scalar and tensor contributions to
$(l+{1\over 2})^2\langle |a_{lm}|^2\rangle$ are approximately constant.
The contribution of scalar perturbations to
$(l+{1\over 2})^2\langle |a_{lm}|^2\rangle$ begins
to decrease for $l\sim 150$ because the scalar
contribution to these multipoles is
dominated by modes that entered the horizon before matter
domination (and hence are suppressed by the
transfer function).  The contribution of tensor modes to
$(l+{1\over 2})^2\langle |a_{lm}|^2\rangle$ begins
to decrease for $l\sim 30$ because the tensor contribution
to these multipoles is
dominated by modes that entered the horizon before decoupling
(and hence decayed as $R^{-1}$ until decoupling).
Figure 10 shows the contribution of scalar and tensor
perturbations to the CBR anisotropy multipole amplitudes
(and includes both the tensor and scalar transfer functions);
the expected variance in the CBR multipoles is given by
the sum of the scalar and tensor contributions.

\begin{figure}
\vspace{4.5in}
\caption[alm]{Scalar and tensor contributions to the
CBR multipole moments:  $l(l+1)\langle |a_{lm}|^2\rangle/6
\langle |a_{2m}|^2\rangle$ for the scalar and $l(l+{1\over 2})
\langle |a_{lm}|^2\rangle /5\langle |a_{2m}|^2\rangle$ for
the tensor with $n-1 = n_T =0$, $z_{\rm DEC} = 1100$,
and $h=0.5$ (from \cite{tl}).  (The tensor angular power spectrum
falls off for $l\sim 30$.)  Scale invariance manifests
itself in the constancy of the angular power spectra for $l\la 100$.
Note, only the Sachs-Wolfe
contribution is shown; for scalar perturbations other
effects become dominant for $l\ga 100$ and the spectrum
rises to a ``Doppler peak'' at around $l\sim 200$, cf. Fig~3.}
\end{figure}

\subsection {Worked examples}

In this Section I apply the formalism developed
in the two previous sections to four specific models.
So that I can, where appropriate, solve numerically
for model parameters, I will:  (1) Assume that
the astrophysically interesting scales crossed
outside the horizon 50 e-folds before the end
of inflation; and (2) Use the COBE DMR quadrupole measurement,
$\langle (\Delta T )_{Q}^2 \rangle^{1/2}
\approx 20\pm 2\mu$K \cite{DMR,bsw}, to normalize
the scalar perturbations; using Eq. (\ref{eq:quadanisotropy}) this implies
\begin{equation}
V_{50} \approx 2.3\times 10^{-11} \,\mpl^4\,x_{50}^2 .
\end{equation} 
Of course it is entirely possible that
a significant portion of the quadrupole anisotropy is
due to tensor-mode perturbations, in which case this normalization
must be reduced by a factor of $(1+T/S)^{-1}$.
And, it is straightforward
to change ``50'' to the number appropriate to a
specific model, or to normalize the perturbations another way.

Before going on let us use the COBE DMR quadrupole anisotropy to bound the
tensor contribution to the quadrupole anisotropy
and thereby the energy density that drives inflation:
\begin{equation}
V_{50}\la 7\times 10^{-11}\mpl^4.
\end{equation}
Thus, the upper limit to the tensor contribution to the CBR quadrupole
implies that the vacuum energy that drives inflation
must be much less than the Planck energy density,
indicating that the final 50 or so e-foldings of inflation,
which is the relevant part of inflation for us, is not a
quantum-gravitational phenomenon.  Of course, inflation
could last far longer than 50 e-foldings and during
the earliest part of inflation the energy density could be
Planckian (this is the point of view advocated by Linde in
his chaotic inflation model \cite{chaotic}).

\subsubsection {Exponential potentials}
 
There are a class of models that can be described in terms
of an exponential potential,
\begin{equation} 
V(\phi ) = V_0\exp (-\beta\phi /\mpl ). 
\end{equation} 
This type of potential was first invoked in the 
context of power-law inflation \cite{powerlaw}, and has 
recently received renewed interest in the context 
of extended inflation \cite{extended}.  In the simplest 
model of extended, or first-order, inflation, 
that based upon the Brans-Dicke-Jordan theory 
of gravity \cite{extended}, $\beta$ is related to the Brans-Dicke
parameter:  $\beta^2 = 64\pi /(2\omega +3)$. 
 
For such a potential the slow-roll conditions are 
satisfied provided that $\beta^2 \la 24\pi$; 
thus inflation does not end until 
the potential changes shape, or in the case of
extended inflation, until the phase transition 
takes place.  In either case we can relate
$\phi_{50}$ to $\phi_{\rm end}$,
\begin{equation} 
N(\phi_{50} ) = 50 = {8\pi\over \mpl^2}\int_{\phi_{50}}^{\phi_{\rm end}} 
{Vd\phi \over -V^\prime};\qquad \Rightarrow\ \ 
\phi_{50} = \phi_{\rm end} - 50\beta / 8\pi . 
\end{equation} 
Since $\phi_{\rm end}$ is in effect arbitrary,
the overall normalization of the potential is
irrelevant.  The two other parameters,
$x_{50}$ and $x_{50}^\prime$, are easy to compute: 
\begin{equation} 
x_{50} = -\beta ; \qquad x_{50}^\prime = 0. 
\end{equation} 
Using the COBE DMR normalization, we can relate
$V_{50}$ and $\beta$: 
\begin{equation} 
V_{50} = 2.3 \times 10^{-11}\,\mpl^4 \beta^2.
\end{equation} 
Further, we can compute $q$, $\alpha_S$, $\alpha_T$, 
and $T/S$: 
\begin{equation} 
q = 16\pi /\beta^2; \qquad T/S = 0.28\beta^2;\qquad 
\alpha_T =\alpha_S = 1/(q-1)\simeq \beta^2/16\pi . 
\end{equation} 
Note, for the exponential potential, $q$, $\alpha_T= 
\alpha_S$ are independent of epoch.  In the case of
extended inflation, $\alpha_S =\alpha_T =4/(2\omega +3)$;
since $\omega$ must be less than about 20 \cite{bigbubbles},
this implies significant tilt:  $\alpha_S=\alpha_T \ga 0.1$.
 
\subsubsection{Chaotic inflation}
 
The simplest chaotic inflation models are based upon potentials of the form:
\begin{equation} 
V(\phi ) = a\phi^b; 
\end{equation} 
$b=4$ corresponds to Linde's original model 
of chaotic inflation and $a$ is dimensionless \cite{chaotic},
and $b=2$ is a model based upon a massive scalar field
and $m^2 = 2a$ \cite{massive}.  In these models $\phi$ is
initially displaced from $\phi = 0$, and inflation
occurs as $\phi$ slowly rolls to the origin. 
The value of $\phi_{\rm end}$ 
is easily found:  $\phi_{\rm end}^2 = b(b-1)\mpl^2/24\pi$, and
\begin{eqnarray}
N(\phi_{50})=50 & =  & {8\pi \over \mpl^2}\int_{\phi_{\rm end}}^{\phi_{50}}
{Vd\phi \over V^\prime} ;\\
& \Rightarrow & \ \
\phi_{50}^2/\mpl^2  =  50b/4\pi + b^2/48\pi \simeq 50b/4\pi ;
\end{eqnarray}
the value of $\phi_{50}$ is a few times the Planck mass. 

For purposes of illustration consider $b=4$; $\phi_{\rm end}
=\mpl /\sqrt{2\pi} \simeq 0.4\mpl$, $\phi_{50} \simeq
4\mpl$, $\phi_{46} \simeq 3.84\mpl$,
and $\phi_{54}\simeq 4.16\mpl$.  In order to have sufficient
inflation the initial value of $\phi$ must exceed about
$4.2\mpl$; inflation ends when $\phi \approx 0.4\mpl$; and
the scales of astrophysical interest cross outside the horizon
over an interval $\Delta \phi \simeq 0.3\mpl$.

The values of the potential, its steepness, and the
change in steepness are easily found,
\begin{equation} 
V_{50} = a\,\mpl^b\,\left({50b\over 4\pi}\right)^{b/2}; 
\qquad x_{50} = \sqrt{4\pi b\over 50}; \qquad 
\mpl x_{50}^\prime = {-4\pi \over 50} ;
\end{equation}
\begin{equation}
q_{50} = 200/b; \qquad
T/S = 0.07b; \qquad \alpha_T\simeq b/200; \qquad 
\alpha_S = \alpha_T + 0.01 . 
\end{equation} 
Unless $b$ is very large, scalar perturbations dominate 
tensor perturbations \cite{star}, $\alpha_T$, $\alpha_S$ are very
small, and $q$ is very large.  Further, when $\alpha_T$, 
$\alpha_S$ become significant, they are equal. 
Using the COBE DMR normalization we find: 
\begin{equation}
a = 2.3 \times 10^{-11} b^{1-b/2} (4\pi /50)^{b/2+1}\,\mpl^{4-b}.
\end{equation} 
For the two special cases of interest: $b=4$, $a=9\times 10^{-14}$;
and $b=2$, $m^2 \equiv 2a = 3\times 10^{-12}\mpl^2$.
 
\subsubsection {New inflation}
 
These models entail 
a very flat potential where the scalar field rolls from 
$\phi \approx 0$ to the minimum of the potential at $\phi
=\sigma$.  The original models of slow-rollover inflation \cite{new}
were based upon potentials of the Coleman-Weinberg form
\begin{equation} 
V(\phi ) = B\sigma^4/2 +B\phi^4\left[ \ln (\phi^2/\sigma^2) 
-{1\over 2} \right] ;
\end{equation} 
where $B$ is a very small dimensionless 
coupling constant.  Other very flat potentials also work (e.g., 
$V = V_0 - \alpha\phi^4 +\beta \phi^6$ \cite{st}).  As before
we first solve for $\phi_{50}$:
\begin{equation} 
N(\phi_{50}) = 50 = {8\pi\over \mpl^2}\int_{\phi_{\rm end}}^{\phi_{50}} 
{Vd\phi\over V^\prime};\qquad \Rightarrow \ \ 
\phi_{50}^2 = {\pi \sigma^4\over 100 |\ln (\phi_{50}^2/\sigma^2)|\mpl^2}; 
\end{equation} 
where the precise value of $\phi_{\rm end}$ is not relevant, 
only the fact that it is much larger than $\phi_{50}$. 
Provided that $\sigma \la \mpl$, both $\phi_{50}$ and 
$\phi_{\rm end}$ are much less than $\sigma$; we then find
\begin{equation} 
V_{50}\simeq B\sigma^4/2; \qquad x_{50} \simeq - {(\pi /25)^{3/2}
\over \sqrt{|\ln (\phi_{50}^2/\sigma^2|) }} \left( {\sigma\over \mpl} 
\right)^2 \ll 1;
\end{equation}
\begin{equation}
\mpl x_{50}^\prime \simeq -24\pi /100;\qquad
q_{50} \simeq {2.5\times 10^5 |\ln (\phi_{50}^2/\sigma^2)|\over \pi^2} 
\left({\mpl\over \sigma}\right)^4 \gg 1;
\end{equation}
\begin{equation}
\alpha_T \simeq {1\over q_{50}} \ll 1;\qquad
\alpha_S = \alpha_T + 0.03;\qquad
{T\over S} \simeq {6\times 10^{-4}\over |\ln (\phi_{50}^2/\sigma^2)|}\left(
{\sigma\over \mpl}\right)^4. 
\end{equation} 
Provided that $\sigma \la \mpl$, $x_{50}$ is very small; this means that
$q$ is very large, gravity-waves
and density perturbations are very nearly scale invariant, 
and $T/S$ is small.  Finally, using the COBE DMR normalization, 
we can determine the dimensionless coupling constant $B$: 
\begin{equation} 
B \simeq 9\times 10^{-14}/|\ln (\phi_{50}^2/\sigma^2)|
\approx 4\times 10^{-15}.
\end{equation} 
 
\subsubsection{Natural inflation}
 
This model is based upon a potential of the form \cite{unnatural}
\begin{equation} 
V(\phi ) = \Lambda^4 \left[ 1+\cos (\phi /f ) \right].
\end{equation} 
The flatness of the potential (and requisite small 
couplings) arise because the $\phi$ particle is a pseudo-Nambu-Goldstone 
boson ($f$ is the scale of spontaneous symmetry breaking 
and $\Lambda$ is the scale of explicit symmetry breaking; in
the limit that $\Lambda \rightarrow 0$ the $\phi$ particle 
is a massless Nambu-Goldstone boson).   It is a simple 
matter to show that $\phi_{\rm end}$ is of the order of $\pi f$. 
 
This potential is difficult to analyze in 
general; however, there are two limiting regimes:
(i) $f\gg \mpl$; and (ii) $f\la \mpl$ \cite{st}.  In the first
regime, the 50 or so relevant e-folds take place close 
to the minimum of the potential, $\sigma = \pi f$, and 
inflation can be analyzed by expanding the potential about $\phi=\sigma$,
\begin{equation} 
V(\psi ) \simeq {m^2\psi^2}/2 ;
\end{equation} 
\begin{equation} 
m^2 = \Lambda^4 /f^2; \qquad \psi = \phi -\sigma . 
\end{equation} 
In this regime natural inflation is equivalent to chaotic
inflation with $m^2 =\Lambda^4 /f^2 \simeq 3\times 10^{-12}\mpl^2$.
 
In the second regime, $f\la \mpl$, inflation takes 
place when $\phi \la \pi f$, so that we can make the 
following approximations:  $V \simeq 2\Lambda^4$ 
and $V^\prime = - \Lambda^4\phi /f^2$.  Taking 
$\phi_{\rm end} \sim \pi f$, we can solve for $N(\phi )$:
\begin{equation} 
N(\phi ) = {8\pi \over \mpl^2}\int_\phi^{\pi f} {Vd\phi \over -V^\prime} 
\simeq {16\pi \mpl^2\over f^2} \ln (\pi f/\phi ) ; 
\end{equation} 
from which it is clear that achieving 50 e-folds of 
inflation places a lower bound to $f$, very 
roughly $f\ga \mpl /3$ \cite{st,unnatural}.

Now we can solve for $\phi_{50}$, $V_{50}$, $x_{50}$,
and $x_{50}^\prime$: 
\begin{equation} 
\phi_{50}/\pi f \simeq \exp (-50 \mpl^2/16\pi f^2)\la {\cal O}(0.1); 
\qquad V_{50} \simeq 2\Lambda^4; 
\end{equation} 
\begin{equation} 
x_{50} \simeq {1\over 2}\,{\mpl\over f}\,{\phi_{50}\over f} 
\la {\cal O}(0.1) ;\qquad 
x_{50}^\prime \simeq - {1\over 2}\,\left( {\mpl\over f}\right)^2. 
\end{equation} 
Using the COBE DMR normalization, we can
relate $\Lambda$ to $f/\mpl$: 
\begin{equation} 
\Lambda /\mpl = 7\times 10^{-4} \sqrt{\mpl\over f}
\exp (-25\mpl^2/16\pi f^2). 
\end{equation} 
Further, we can solve for $T/S$, $\alpha_T$, and $\alpha_S$: 
\begin{equation} 
{T\over S} \simeq 0.07 \left( {\mpl\over f}\right)^2
\left({\phi_{50} \over f}\right)^2 \la {\cal O}(0.1) ; 
\end{equation} 
\begin{equation} 
\alpha_T = {1\over 16\pi}\,{q_{50}\over q_{50}-1} \left( 
{1\over 4} {\mpl^2\over f^2}{\phi_{50}^2\over f^2} \right) 
\approx  {1\over 64\pi}
\left({\mpl\over f}\right)^2\left({\phi_{50}\over f}\right)^2\ll 0.1; 
\end{equation} 
\begin{equation} 
\alpha_S = {1\over 16\pi}\,{q_{50}\over q_{50}-1} \left( 
{1\over 4}{\mpl^2\over f^2}{\phi_{50}^2\over f^2} + {\mpl^2\over 
f^2}\right) \approx {1\over 16\pi}\left({\mpl\over f}\right)^2; 
\end{equation} 
\begin{equation} 
q_{50} = 64\pi \left({f\over \mpl}\right)^2\left( 
{f\over \phi_{50}}\right)^2  \gg 1. 
\end{equation} 

Regime (ii) provides the exception
to the rule that $\alpha_S\approx\alpha_T$ and large
$\alpha_S$ implies large $T/S$.  For example, taking 
$f=\mpl /2$, we find: 
\begin{equation} 
\phi_{50}/f \sim 0.06; \qquad x_{50} \sim 0.06; \qquad 
x_{50}^\prime = - 2; \qquad q_{50} \sim 10^4;
\end{equation} 
\begin{equation} 
\alpha_T \sim 10^{-4};\qquad \alpha_S \sim 0.08;\qquad T/S \sim 10^{-3}. 
\end{equation} 
The gravitational-wave perturbations are very nearly scale 
invariant, while the density perturbations deviate 
significantly from scale invariance.  I note that
regime (ii), i.e., $f \la \mpl$, occupies only a tiny fraction of parameter
space because $f$ must
be greater than about $\mpl /3$ to achieve sufficient
inflation; further, regime (ii) is ``fine tuned'' and
``unnatural'' in the sense that the required value of $\Lambda$ is
exponentially sensitive to the value of $f/\mpl$.

Finally, I note that the results for regime (ii)
apply to any inflationary model whose Taylor expansion
in the inflationary region is similar; e.g., $V(\phi )=
-m^2\phi^2 + \lambda\phi^4$, which was originally analyzed
in Ref.~\cite{st}.

\subsubsection{Lessons}

To summarize the general features of our results.
In all examples the deviations from scale invariance
enhance perturbations on large scales.  The only
potentials that have significant deviations from
scale invariance are either very steep or have rapidly
changing steepness.  In the former case, both the
scalar and tensor perturbations are tilted by a
similar amount; in the latter case, only the scalar
perturbations are tilted.

For ``steep'' potentials,
the expansion rate is ``slow,'' i.e., $q_{50}$ close to unity,
the gravity-wave contribution to the CBR quadrupole anisotropy
becomes comparable to, or greater than, that of density
perturbations, and both scalar and tensor
perturbations are tilted significantly.
For flat potentials, i.e., small $x_{50}$,
the expansion rate is ``fast,'' i.e., $q_{50} \gg 1$,
the gravity-wave contribution to the CBR quadrupole
is much smaller than that of density perturbations, and unless
the steepness of the potential changes significantly,
large $x_{50}^\prime$, both spectra very nearly scale invariant;
if the steepness of the potential changes rapidly,
the spectrum of scalar perturbations can be tilted significantly.
The models that permit significant deviations from scale
invariance involve exponential or low-order polynomial
potentials; the former by virtue of their steepness, the latter
by virtue of the rapid variation of their steepness.
Exponential potentials are of interest because they arise
in extended inflation models; potentials with rapidly
steepness include $V(\phi ) =-m^2\phi^2+\lambda\phi^4$
or $\Lambda^4[1+\cos (\phi /f)]$.

Finally, to illustrate how observational data could used
to determine the properties of the inflationary potential
and test the consistency of the inflationary hypothesis,
suppose observations determined the following:
\begin{equation}
(\Delta T)_Q  \simeq 16\mu{\rm K};
\qquad T/S = 0.24; \qquad n = 0.9;
\end{equation}
that is, the COBE DMR quadrupole anisotropy, a four to one ratio
of scalar to tensor contribution to the CBR quadrupole,
and spectral index of 0.9 for the scalar perturbations.
From $T/S$, we determine the steepness of the potential:
$x_{50} \simeq 0.94$.  From the steepness and the
quadrupole anisotropy the value of the potential:
$V_{50}^{1/4}\simeq 2.4\times 10^{16}\GeV$.  From
the spectral index the change in steepness:
$x_{50}^\prime \simeq -0.81/\mpl$.
These data can also be expressed in terms of the
value of the potential and its first two derivatives:
\begin{equation}
V_{50} = 1.4\times 10^{-11}\mpl^4;\qquad
V_{50}^\prime = 1.5\times 10^{-11}\mpl^3;\qquad
V_{50}^{\prime\prime} = 1.0\times 10^{-12}\mpl^2.
\end{equation}
Further, they the lead to the prediction:
$n_T = -0.035$, which, when ``measured,'' can be used
as a consistency check for inflation.

\section{STRUCTURE FORMATION:
CRUCIAL TEST OF INFLATION}

The key to testing inflation is to focus on its robust predictions
and their implications.  Earlier I discussed the prediction of
a flat Universe and its bold implication that most of the matter in Universe
exists in the form of particle dark matter.  Much effort is being
directed at determining the mean density of the Universe and
detecting particle dark matter.

The scale-invariant scalar metric perturbations
lead to CBR anisotropy on angular scales from less
than $1^\circ$ to $90^\circ$
and seed the formation of structure in the Universe.
Together with the nucleosynthesis determination of $\Omega_B$ and
the inflationary prediction of a flat Universe, scale-invariant
density perturbations lead to a very specific scenario for structure
formation; it is known
as cold dark matter because the bulk of the particle dark
matter is comprised of slowly moving particles (e.g., axions or
neutralinos) \cite{cdmrev}.\footnote{The simpler possibility,
that the particle dark matter exists in the form of $30\eV$ or so
neutrinos which is known as hot dark matter, was falsified almost
a decade ago.  Because neutrinos move rapidly, they can diffuse
from high density to low density regions damping perturbations on
small scales.  In hot dark matter large, supercluster-size objects
must form before galaxies, and thus hot dark matter cannot account
for the abundance of galaxies, damped Lyman-$\alpha$ clouds, etc.
that is observed at high redshift.}   A large and rapidly growing number
of observations are being brought to bear in the testing of
cold dark matter, making it the centerpiece of efforts to test inflation.

Finally, there are the scale-invariant tensor perturbations.  They
lead to CBR anisotropy on angular scales from a few degrees to
$90^\circ$ and a spectrum of gravitational waves.  The CBR anisotropy
arising from the tensor perturbations can in principle be separated
from that arising from scalar perturbations.  However, because the sky is
finite, sampling variance sets a fundamental limit:  the tensor
contribution to CBR anisotropy can only be separated from that of
the scalar if $T/S$ is greater than about $0.14$ \cite{knoxmst}.  It is
also possible that the stochastic background of gravitational
waves itself can be directly detected, though it appears that
the LIGO facilities being built will lack the sensitivity and
even space-based interferometery (e.g., LISA) is not a sure bet \cite{tlw}.

Before going on to discuss how cold dark matter models are testing
inflation I want to emphasize the importance
of the tensor perturbations.   The attractiveness of a flat Universe
with scale-invariant density perturbations was appreciated long before
inflation.  Verifying these two predictions of inflation, while
important, will not provide a ``smoking gun.''  The tensor perturbations
are a unique feature of inflation.  Further, they are crucial
to obtaining information about the scalar potential responsible for inflation.

\subsection{Vanilla Cold Dark Matter: almost, but not quite?}

Cold dark matter has often been characterized as a
``no parameter model'' for structure formation; that is
only true in the broad brush:  cold dark matter is characterized
by scale-invariant density perturbations and a matter content
that is almost entirely slowly moving particles.  To make predictions of
the precision needed to match current observations, a more specific
characterization is essential -- precise power-law index of the
spectrum of density perturbations, amplitude of tensor
perturbations, Hubble constant, baryon density,
radiation content of the Universe, possible cosmological constant, and so on.

Historically, the ``standard'' version of
cold dark matter, vanilla cold dark matter
if you will, is :  (1) $\Omega_B \simeq 0.05$
and $\Omega_{\rm CDM} \sim 0.95$; (2) Hubble constant of $50\kms\Mpc^{-1}$;
(3) Precisely scale-invariant density perturbations ($n=1$); and
(4) No contribution of tensor perturbations to CBR anisotropy.
Standard cold dark matter has no other significance than as
a default starting point.
Because it became an ``industry standard''
vanilla cold dark matter provides an interesting
point of comparison -- but that is all!

In cold dark matter models structure forms hierarchically, with small objects
forming first and merging to form larger objects.  Galaxies form
at redshifts of order a few, and rarer objects like QSOs form from higher
than average density peaks earlier.
In general, cold dark matter predicts a Universe that is still
evolving at recent epochs.  $N$-body simulations
are crucial to bridging the gap between theory and observation,
and several groups have carried out large numerical studies of
vanilla cold dark matter \cite{nbody}.

There are a diversity of observations that test cold dark matter; they
include CBR anisotropy and spectral distortions, redshift surveys, pairwise
velocities of galaxies, peculiar velocities,
redshift space distortions, x-ray background,
QSO absorption line systems, cluster studies of all kinds,
studies of evolution (clusters, galaxies, and so on), measurements
of the Hubble constant, and on and on.  I will focus on how these
measurements probe the power spectrum of density perturbations,
emphasizing the role of
CBR-anisotropy measurements and redshift surveys.

Density perturbations on a (comoving) length scale $\lambda$
give rise to CBR anisotropy on an angular scale $\theta \sim
\lambda/H_0^{-1} \sim 1^\circ (\lambda /100h^{-1}\Mpc )$.\footnote{For
reference, perturbations on a length scale of about $1\Mpc$ give
rise to galaxies, on about $10\Mpc$ to clusters, on about $30\Mpc$
to large voids, and on about $100\Mpc$ to the great walls.}
CBR anisotropy has now been detected by more than ten experiments
on angular scales from about $0.5^\circ$ to $90^\circ$, thereby
probing length scales from $30h^{-1}\Mpc$ to $10^4h^{-1}\Mpc$.
The very accurate measurements made by the COBE DMR can be used
to normalize the cold dark matter spectrum (the normalization
scale corresponds to about $20^\circ$).  When this is done, the
other ten or so measurements are in agreement with the predictions
of cold dark matter (see Fig.~1).

The COBE-normalized cold dark matter spectrum can be extrapolated
to the much smaller scales probed by redshift surveys, from about
$1h^{-1}\Mpc$ to $100h^{-1}\Mpc$.  When this is done, there is general
agreement.  However, on closer inspection the COBE-normalized spectrum
seems to predict excess power on these scales (about a factor
of four in the power spectrum; see Fig.~7).  This conclusion
is supported by other observations, e.g., the abundance of rich
clusters and the pairwise velocities of galaxies.  It suggests
that cold dark matter has much of the truth, but perhaps not
all of it \cite{cdmproblems}, and has led to the suggestion that
something needs to be added to the simplest cold dark matter theory.

There is another important challenge facing cold dark matter.
X-ray observations of rich clusters are able to determine the ratio
of hot gas (baryons) to total cluster mass (baryons + CDM) (by a wide
margin, most of the baryons ``seen'' in clusters are in the hot gas).
To be sure there are assumptions and uncertainties; the data at the moment
indicate that this ratio is $(0.04 -0.08) h^{-3/2}$ \cite{gasratio}.
If clusters provide a fair sample of the universal mix of matter,
then this ratio should equal $\Omega_B/(\Omega_B + \Omega_{\rm CDM})
\simeq (0.009-0.022)h^{-2}/ (\Omega_B + \Omega_{\rm CDM})$.  Since
clusters are large objects they should provide a pretty fair sample.
Taking the numbers at face value, cold dark matter is consistent with the
cluster gas fraction provided either:  $\Omega_B + \Omega_{\rm CDM}
= 1$ and $h\sim 0.3$ or $\Omega_B + \Omega_{\rm CDM} \sim 0.3$ and
$h\sim 0.7$.  The cluster baryon problem has yet to be settled,
and is clearly an important test of cold dark matter.
                                                       
Finally, before going on to discuss the variants of cold dark matter now under
consideration, let me add a note of caution.
The comparison of predictions for structure formation
with present-day observations of the
distribution of galaxies is fraught with difficulties.
Theory most accurately predicts ``where the mass is''
(in a statistical sense) and the observations determine where the light is.
Redshift surveys probe present-day inhomogeneity on scales
from around one $\Mpc$ to a few hundred $\Mpc$, scales where
the Universe is nonlinear ($\delta n_{\rm GAL}/n_{\rm GAL}
\ga 1$ on scales $\la 8h^{-1}\Mpc$) and where astrophysical
processes undoubtedly play an important role
(e.g., star formation determines where and when
``mass lights up,'' the explosive release of energy in supernovae
can move matter around and influence subsequent star formation,
and so on).  The distance to a galaxy is
determined through Hubble's law ($d = H_0^{-1} z$) by measuring a redshift;
peculiar velocities induced by the lumpy distribution
of matter are significant and prevent a direct determination
of the actual distance.  There are the intrinsic limitations
of the surveys themselves:  they are flux not volume limited (brighter
objects are seen to greater distances and vice versa) and relatively
small (e.g., the CfA slices of the Universe survey contains only
about $10^4$ galaxies and extends to a redshift of about $z\sim 0.03$).
Last but not least are the numerical
simulations which link theory and observation;
they are limited in dynamical range (about a factor
of 100 in length scale) and in microphysics (in the largest simulations
only gravity, and in others only a gross approximation to the effects of
hydrodynamics/thermodynamics).  Perhaps it would be prudent to withhold
judgment on vanilla cold dark matter for the moment and resist
the urge to modify it---but that wouldn't be as much fun!

\subsection{The many flavors of cold dark matter}

The spectrum of density perturbations today depends
not only upon the primeval spectrum (and the normalization
on large scales provided by COBE), but also upon the energy content
of the Universe.  While the fluctuations in the gravitational potential
were initially (approximately) scale invariant,
the Universe evolved from an early radiation-dominated phase
to a matter-dominated phase which imposes a characteristic scale
on the spectrum of density perturbations seen today;
that scale is determined by the energy
content of the Universe, $k_{\rm EQ}\sim 10^{-1}h\Mpc^{-1}
\,(\Omega_{\rm matter}h/\sqrt{g_*})$
($g_*$ counts the relativistic degrees of freedom,
$\Omega_{\rm matter} = \Omega_B +\Omega_{\rm CDM}$).
In addition, if some of the nonbaryonic dark
matter is neutrinos, they reduce power on small
scales somewhat through freestreaming (see Fig.~7).
With this in mind, let me discuss
the variants of cold dark matter that have been proposed to
improve its agreement with observations.

\begin{enumerate}

\item {\bf Low Hubble Constant + cold dark matter (LHC CDM) \cite{lhc}.}
Remarkably, simply lowering the Hubble constant to around
$30\kms\Mpc^{-1}$ solves all the problems
of cold dark matter.  Recall, the critical density $\rho_{\rm crit}
\propto H_0^2$; lowering $H_0$ lowers the matter density and
has precisely the desired effect.  It has two other added benefits:
the expansion age of the Universe is comfortably
consistent with the ages of the
oldest stars and the baryon fraction is raised to a value that is
consistent with that measured in x-ray clusters.  Needless to say, such a
small value for the Hubble constant flies in the face of current
observations \cite{distance2,freedmanetal};
further, it illustrates that the problems of
cold dark matter get even worse for the larger values of $H_0$
that are favored by recent observations.

\item {\bf Hot + cold dark matter ($\nu$CDM) \cite{chdm}.}
Adding a small amount of
hot dark matter can suppress density perturbations on small scales;
adding too much leads back to the longstanding
problems of hot dark matter.  Retaining enough power
on very small scales to produce damped Lyman-$\alpha$ systems at
high redshift limits $\Omega_\nu$ to less than about
20\%, corresponding to about ``$5\eV$ worth
of neutrinos'' (i.e., one species of mass $5\eV$, or two species
of mass $2.5\eV$, and so on).  This admixture of hot dark matter
rejuvenates cold dark matter provided the Hubble constant is not
too large, $H_0\la 55 \kms\Mpc^{-1}$; in fact, a Hubble constant
of closer to $45\kms\Mpc^{-1}$ is preferred.

\item {\bf Cosmological constant + cold dark matter ($\Lambda$CDM) \cite{lcdm}.}
(A cosmological constant corresponds to a uniform energy density,
or vacuum energy.)  Shifting 50\% to 70\% of the critical density
to a cosmological constant lowers the matter density
and has the same beneficial effect as a low Hubble constant.
In fact, a Hubble constant as large as $80\kms\Mpc^{-1}$
can be accommodated.  In addition,
the cosmological constant allows the age problem to solved
even if the Hubble constant is large, addresses the fact
that few measurements of the mean mass density give a value as large as
the critical density (most measurements of the mass density
are insensitive to a uniform component), and allows the
baryon fraction of matter to be larger, which
alleviates the cluster baryon problem.   Not everything is rosy;
cosmologists have invoked a cosmological constant twice before to solve
their problems (Einstein to obtain a static universe and
Bondi, Gold, and Hoyle to solve the earlier age crisis when
$H_0$ was thought to be $250\kms\Mpc^{-1}$).  Further, particle
physicists can still not explain why the energy of the
vacuum is not at least 50 (if not 120) orders of magnitude larger than
the present critical density, and expect that when the problem is
solved the answer will be zero.

\item {\bf Extra relativistic particles + cold
dark matter ($\tau$CDM) \cite{taucdm}.}
Raising the level of radiation has the same beneficial effect
as lowering the matter density.  In the standard cosmology the
radiation content consists of photons + three (undetected)
cosmic seas of neutrinos (corresponding to $g_* \simeq
3.36$).   While we have no direct determination
of the radiation beyond that in the CBR, there are
at least two problems:  What are the additional relativistic
particles? and Can additional radiation be added without
upsetting the successful predictions
of primordial nucleosynthesis which depend critically upon the
energy density of relativistic particles?  The simplest way around
these problems is an unstable tau neutrino (mass anywhere
between a few keV and a few MeV) whose decays produce
the radiation.   This fix can tolerate a larger Hubble constant,
though at the expense of more radiation.

\item {\bf Tilted cold dark matter (TCDM) \cite{tcdm}.}
While the spectrum of density
perturbations in most models of inflation is very nearly scale invariant,
there are models where the deviations are significant ($n\approx 0.8$)
which leads to smaller fluctuations on small scales.  Further,
if gravity waves account for a significant
part of the CBR anisotropy, the level of density perturbations can be
lowered even more.  A combination of tilt and gravity waves can solve the
problem of too much power on small scales, but seems to lead to
too little power on intermediate and very small scales.

\end{enumerate}

These possibilities represent different approaches to improving the
concordance of CDM.  They also represent well motivated
modifications to the standard cosmology in their
own right.  It has always been
appreciated that the inflationary spectrum of density perturbations
was not exactly scale invariant \cite{st} and that the Hubble constant
was unlikely to be exactly $50\kms \Mpc$.  Neutrinos exist; they are
expected to have mass;
there is even some experimental data that indicates they do have
mass \cite{numass}.  If the Hubble constant is as large
as $70\kms\Mpc^{-1}$ to $80\kms\Mpc^{-1}$ a cosmological constant
seems inescapable based upon the age of the Universe alone.
There is no data precludes more radiation than in the standard cosmology.
In fact, these modifications to vanilla cold dark matter are so
well motivated that one should probably also consider
combinations; e.g., lesser tilt and $h=0.45$ and so on \cite{stp}.

In evaluating these better fit models, one should keep the words
of Francis Crick in mind (loosely paraphrased):  A model that fits
all the data at a given time is necessarily wrong, because at any given
time not all the data are correct(!).  $\Lambda$CDM provides an
interesting/confusing example.  When I discussed it in 1990,
I called it the best-fit Universe, and quoting Crick, I said that
$\Lambda$CDM was certain to fall by the wayside \cite{lcdm0}.
In 1995, it is still the best-fit model \cite{lcdm1}.

\subsection{Reconstruction}

If inflation and the cold dark matter theory is shown to be correct,
then a window to the very early Universe ($t\sim 10^{-34}\sec$) will
have been opened.  While it is certainly premature to jump to this
conclusion, I would like to illustrate one example of what one could
hope to learn.  As mentioned earlier, the spectra and amplitudes
of the the tensor and scalar metric perturbations predicted by
inflation depend upon the underlying model, to be specific, the
shape of the inflationary scalar-field potential.
If one can measure the power-law
index of the scalar spectrum and the amplitudes of the scalar
and tensor spectra, one can recover the value of the potential
and its first two derivatives around the point on the potential
where inflation took place \cite{reconstruct}:
\begin{eqnarray}
V & = & 1.65 T\, {m_{\rm Pl}}^4  , \\
V^\prime & = & \pm \sqrt{8\pi r \over 7}\, V/{m_{\rm Pl}} , \\
V^{\prime\prime} & = & 4\pi \left[ (n-1) + {3\over 7} r \right]\,
V /{m_{\rm Pl}}^2 ,
\end{eqnarray}
where $r\equiv T/S$, a prime indicates derivative
with respect to $\phi$, $\mpl = 1.22\times 10^{19}\GeV$ is
the Planck energy, and the sign of $V^\prime$ is
indeterminate.  In addition, if the tensor spectral index
can be measured a consistency relation, $n_T = -r/7$,
can be used to further test inflation.
Reconstruction of the inflationary scalar potential would
shed light both on inflation as well as physics at energies of the
order of $10^{15}\GeV$.  (If $\Lambda \not= 0$, these expressions are
modified \cite{lrecon}.)

\section{The Future}

The stakes for cosmology are high:  if correct, inflation/cold dark matter
represents a major extension of the big bang and our understanding
of the Universe.  Further, it will shed
light on the fundamental physics at energies of order $10^{15}\GeV$.

What are the crucial tests and when will they be carried out?
Because of the many measurements/observations
that can have significant impact, I believe the answer to when
is sooner rather than later.  The list of pivotal observations is long:
CBR anisotropy, large redshift surveys (e.g., the Sloan Digital
Sky Survey will have $10^6$ redshifts), direct searches
for nonbaryonic in our neighborhood (both for axions and neutralinos)
and baryonic dark matter (microlensing), x-ray studies of galaxy
clusters, the use of back-lit gas clouds (quasar absorption line systems)
to study the Universe at high redshift, evolution (as
revealed by deep images of the sky taken by the Hubble Space Telescope and
the Keck 10 meter telescope), measurements of both $H_0$ and $q_0$,
mapping of the peculiar velocity field at large redshifts
through the Sunyaev-Zel'dovich effect, dynamical estimates of
the mass density (using weak gravitational lensing, large-scale velocity
fields, and so on), age determinations,
gravitational lensing, searches for supersymmetric
particles (at accelerators) and neutrino oscillations (at accelerators,
solar-neutrino detectors, and other large underground detectors),
searches for high-energy neutrinos from
neutralino annihilations in the sun using large underground detectors,
and on and on.  Let me end by illustrating the interesting
consequences of several possible measurements.

A definitive determination that $H_0$ is greater than $55\kms\Mpc^{-1}$
would falsify all CDM models except that with a cosmological
constant and would certainly give particle theorists
something to think about.  (A definitive determination
that $H_0$ is $75\kms \Mpc^{-1}$ or larger would necessitate a
cosmological constant based upon the age of the Universe
alone, though it should be noted that none of the CDM models
consistent with large-scale structure have an age problem.)  A flat Universe
with a cosmological constant has a very different deceleration
parameter than one dominated by matter, $q_0 =-1.5\Omega_\Lambda + 0.5
\sim -(0.4 - 0.7)$ compared to $q_0 = 0.5$, and this could be settled
by galaxy-number counts, quasar-lensing statistics, or a Hubble
diagram based upon Type Ia supernovae.
The predicted CBR anisotropy on the $0.5^\circ$ scale in $\tau$CDM and LHC CDM
is about 50\% larger than vanilla CDM and about 50\% smaller
in TCDM, which should be easily discernible.   If neutrino-oscillation
experiments were to provide evidence for a neutrino of mass $5\eV$ (or two of
mass $2.5\eV$) $\nu$CDM would seem almost inescapable \cite{numass}.

More CBR measurements are in progress and there should
many interesting results in the next few years.  In the wake
of the success of COBE there are proposals, both in the US and
Europe, for a satellite-borne instrument to map the CBR sky
with a factor of thirty or more better resolution.
A map of the CBR with $0.3^\circ$ resolution could separate
the gravity-wave contribution to
CBR anisotropy and provide evidence for the third robust
prediction of inflation, as well as determining other important
parameters \cite{knoxphd}, e.g., the scalar and tensor indices,
$\Omega_\Lambda$, and even $\Omega_0$ (the position of
the ``Doppler'' peak scales as $\sqrt{\Omega_0}\,$degrees \cite{omega}).

The future in cosmology is very bright:  We have a highly successful
standard model---the hot big-bang; bold ideas for extending it---inflation
and cold dark matter; and a flood of data to test these ideas.

\end{document}